%% file: blobs.tex
\documentclass[usenatbib,usegraphicx,useAMS]{mn2e}
\usepackage{amsmath}
\usepackage{amssymb}

\input{journal_abbr}  
\input{standard_abbr} 
\input{unit_abbr} 

\newcommand{\SgrA}{Sgr A*}

\def\ImageFigPath{LowResImages}

\title{Imaging Bright Spots in the Accretion Flow Near the 
Black Hole Horizon of \SgrA}
\author[Avery E. Broderick \& Abraham Loeb]{Avery
E. Broderick\thanks{E-mail: abroderick@cfa.harvard.edu} \& Abraham
Loeb\thanks{E-mail: aloeb@cfa.harvard.edu}\\ Institute for Theory and
Computation, Harvard-Smithsonian Center for Astrophysics, 60 Garden St., MS
51, Cambridge, MA 02138, USA\\}

\begin{document}
\maketitle

\begin{abstract}
Images from the vicinity of the black hole horizon at the Galactic centre
(\SgrA) could be obtained in the near future with a Very Large Baseline
Array of sub-millimetre telescopes.  The recently observed short-term infrared
and X-ray variability of the emission from \SgrA~implies that the
accretion flow near the black hole is clumpy or unsteady.  We
calculate the appearance of a compact emission region (bright spot) in
a circular orbit around a spinning black hole as a function of orbital
radius and orientation. We find that the mass and spin of the black
hole can be extracted from their generic signatures on the spot image
as well as on the lightcurves of its observed flux and
polarisation. The strong-field distortion remains noticeable even when
the spot image is smoothed over the expected $\sim 20\,\muas$ resolution
of future sub-millimetre observations.
\end{abstract}


\section{Introduction} \label{I}
Despite its successes, there has not yet been a satisfactory test
of general relativity in the strong gravity limit.  By their very nature,
studies of black holes are likely to provide the best opportunity for
constraining strong field relativity.  Unfortunately, current attempts to
do this rely upon modelling the accretion flow and are thus indirect
\citep[see,
\eg,][]{Nara-Heyl:02,Tana-etal:95,Reyn-Nowa:03,Pari-Brom-Mill:01}.

Of these, the most direct method involves observations of line-like
features in the X-ray spectra of black hole candidates, typically
interpreted as the Fe K$\alpha$ fluorescence line, broadened as a result of
the Keplerian motion of the disk and frame dragging due to the rotation of
the black hole \citep[see, \eg,][]{Reyn-Nowa:03,Pari-Brom-Mill:01}.  The
lack of a correlation between the variability in the soft X-ray
continuum and the line emission implies that the simplest model used
for the interpretation of the observations is incomplete \citep[see,
\eg,][]{Weav-Gelb-Yaqo:01,Wang-Wang-Zhou:01,Wang-Zhou-Xu-Wang:99,Chia_etal:00,Lee_etal:00},
though attempts to rectify this situation with the inclusion of
gravitational lensing have been made \citep{Matt-Fabi-Reyn:97}.  The case
is further complicated by the existence of alternative interpretations
\citep[see, \eg,][]{Elvi:00,You-Liu-Chen-Chen-Zhan:03}.

General relativistic effects can also play a substantial role in the
polarisation properties of Thomson thick disks \citep[see,
\eg][]{Conn-Star:80,Laor-Netz-Pira:90,Bao-Hadr-Wiit-Xion:97}.  Therefore,
detailed spectropolarimetric observations may shed light on both
the physics of the accretion flow and the curvature of space-time.
However, these necessarily require an accretion model and consequently
suffer from considerable uncertainties associated with the accretion
physics.

In contrast, it may be possible to directly image a black hole, and
thus measure the space-time curvature down to the {\em photon orbit},
the radius at which photons execute a circular orbit,
$3GM/c^2$ for a Schwarzschild black hole and down to $GM/c^2$
for a maximally
rotating Kerr black hole\footnote{Note that the coincidence between the
photon orbit and the horizon in the maximally rotating Kerr case is an
artifact of Boyer-Lindquist coordinates.  At all values of black hole spin, a
non-vanishing radial proper distance exists between the horizon and
the photon orbit \citep[see, \eg,][]{Chan:92}.}
\citep{Falc-Meli-Agol:00}. The black hole in the Galactic centre
provides the best candidate for such an observation as it possesses
the largest apparent size on the sky, with $GM/c^2$ corresponding to an
angular scale of $\sim5\,\muas$.  Within the next decade it is
expected that a Very Large Baseline Array (VLBA) at sub-millimetre
wavelengths (where scattering is no longer the
limiting factor) will exist which is expected to provide $\sim20\,\muas$
resolution \citep{Gree:05,Miyo_etal:04}.

Earlier theoretical work has focused on optically thin, azimuthally
symmetric accretion flows, which are generically expected to show a
shadow around the black hole
\citep[\eg,][]{Falc-Meli-Agol:00,Taka:05,Taka:04,Beck-Done:05}.  However,
numerical general-relativistic magnetohydrodynamic simulations of
accretion flows suggest that
the region near the innermost stable circular orbit (ISCO) may be strongly
inhomogeneous \citep[see, \eg,][]{DeVi-Hawl-Krol:03}.  Indeed, recent
observations of \SgrA~in the infrared and X-ray bands revealed flaring
activity on short time scales
\citep{Ghez_etal:04,Ecka_etal:04,Genz_etal:03,Porq_etal:03,Asch_etal:04,Gold_etal:03,Baga_etal:01}
indicating strong inhomogeneities in the
emission close to the black hole horizon. Computations of the light curves
of inhomogeneous accretion flows have been performed in the context of
quasars and X-ray binaries more generally
\citep{Bao-Wiit-Hadr:98,Bao-Hadr-Wiit-Xion:97}, but without providing
images to be compared with future observations and without studying the
dependence of the light curve on the black hole parameters.

In this paper, the unpolarised and polarised light curves and
images are computed for an optically thick emitting sphere at a number
of radii, viewing angles, and black hole spins.  While this may
strictly apply for a star orbiting a supermassive black hole, it should
be understood that our treatment is a proxy for any mechanism which
enhances the emission in a compact region of space (\eg, due to
reconnection events, over densities in mass or magnetic field
strength).

The method by which rays are traced, the radiative transfer performed, and
the optically thick sphere is modelled are discussed in section \ref{CM}.
Section \ref{LC} compares the expected light curves (both polarised and
unpolarised) for a number of orbits, viewing angles, and black holes spins.
Images and centroid motions are presented in sections \ref{Im} and
\ref{Ce}, respectively.  Finally, concluding remarks are summarised in
section \ref{C}.

In what follows, the metric signature is taken to be \mbox{$(-+++)$},
and geometrised unites are used ($G=c=1$).

\section{Computational Methods} \label{CM}
Due to the complexities introduced by the Kerr metric, in the strong field
limit, images of the accretion flow are most readily obtained via numerical
methods.  This computation may be succinctly segregated into the problems
of tracing null geodesics, performing the radiative transfer along these
rays, and modelling the inhomogeneity (in this case an optically thick
sphere). The first two of these are treated in subsection \ref{CM:RTRT},
and the third in subsection \ref{CM:OTS}.

\subsection{Ray Tracing \& Radiative Transfer} \label{CM:RTRT}

The light rays are traced in curved space following the scheme of
\citet{Brod-Blan:03} (for which tracing null geodesics is a limiting case),
as well as the radiative transfer method of \citet{Brod-Blan:04}.  Below,
only the essential aspects are summarised.

Null geodesics are constructed by integrating the equations
\begin{align}
\frac{d x^\mu}{d\lambda} &= f(r) k^\mu \nonumber\\
\frac{d k_\mu}{d\lambda} &= - f(r) \left( \frac{1}{2} \frac{\partial k^\nu
  k_\nu}{\partial x^\mu} \right)_{k_\alpha} \,,
\end{align}
where the partial differentiation is taken holding the covariant
components of the wave four-vector, $k_\mu$, constant, and
\begin{equation}
f(r) = r^2 \sqrt{1 - \frac{r_h}{r}}\,,
\end{equation}
(where $r_h$ is the horizon radius)
simply reparameterises the affine parameter, $\lambda$, along the ray to
avoid singular behaviour near the horizon.  An explicit demonstration that
$x^\mu(\lambda)$ reproduce the null geodesics can be found in
\citet{Brod-Blan:03}.

Typically, polarised radiative transfer is performed using the Stokes
parameters.  However, since these are not Lorentz scalars, in relativistic
environments they are not ideal.  In contrast, the photon distribution function
($\propto I_\nu/\nu^3$) is a Lorentz scalar, and thus much simpler to
evolve along the null geodesics.  As shown in \citet{Brod-Blan:04}, it is
possible to define analogues of the photon distribution function for the rest
of the Stokes parameters using an orthonormal tetrad propagated along the
null geodesics.

While the numerical scheme utilised here is not optimized for this
particular problem, it has the virtue of already being implemented and
tested in the context of imaging accreting black holes.

\subsection{Optically Thick Sphere} \label{CM:OTS}
An inhomogeneity in the accretion flow is modelled as an
optically thick sphere orbiting around the central black hole.  The
surface of the inhomogeneity is given implicitly by
\begin{equation}
\Delta r^\mu \Delta r_\mu + \left( u_S^\mu \Delta r_\mu \right)^2 = R_S^2
\label{sphere}
\end{equation}
where $\Delta r^\mu \equiv r^\mu - r_S^\mu$ is the displacement from the
sphere centre, located at $r_S^\mu$, moving with velocity $u_S^\mu$, and
$R_S$ is the radius of the sphere.  The second term accounts for the length
contraction\footnote{Despite the length contraction, it is long been known
that in the far field, relativistic aberration combined with time of flight
effects result in the relativistically moving sphere appearing rotated but
otherwise unchanged \citep{Penr:58,Terr:59}.  This may indeed be seen
explicitly in the images presented in section \ref{Im}.}.  In the limit
that $\Delta r^\mu \Delta r_\mu \ll r$ this definition indeed produces
a sphere in the comoving frame.  For $\Delta r^\mu \Delta r_\mu \sim
r$ the surface defined by equation (\ref{sphere}) necessarily departs from sphericity (since the $\Delta r^\mu
\Delta r_\mu$ is only the differential line element), and thus in
reality is only
quasi-spherical.  However, considering the level of approximation inherent
in treating the inhomogeneity as a sphere in the first place, the added
simplicity outweighs a more detailed effort to produce a true sphere.  Here
a value of $R_S = 1.5 M$ is used, which is sufficiently compact to reveal the
strong field effects of interest in this study.

Since the sphere is an extended body, it cannot have a uniform velocity if
it is to remain coherent.  Here all points upon the surface of the sphere
are taken to have the same angular velocity as measured by the
Boyer-Lindquist observer, \ie, the same $u_S^\phi/u_S^t$, ensuring that all
parts of the sphere execute an orbit in the same amount of observer time.
An explicit proof that this ensures that the sphere does not shear can be
found in \citet{Brod-Blan:04}.  The velocity of the centre of the sphere
was taken to be that associated with the stable circular orbit at
$r_S^\mu$.

Images are produced by tracing a bundle of parallel rays back from a
screen far from the black hole.  Rays that impinged upon the sphere were
given an intensity (subsequently transformed into a photon
distribution function) governed by the Rayleigh-Jeans law,
\begin{equation}
I_\nu = 2\frac{\nu^2}{c^2} k T\,,
\end{equation}
where $\nu = -u_\mu k^\mu/2\pi$ and the temperature was assumed to be
proportional to its virial value:
\begin{equation}
T \propto - \left( u_t - \frac{1}{\sqrt{-g^{tt}}} \right)\,,
\end{equation}
where $g^{\mu\nu}$ are the contravariant components of the 
metric.

Polarisation measurements place an additional observational constraint upon
the emission and propagation physics.  Because the polarisation is
parallel--propagated along the ray, it is especially sensitive to the
effects of strong gravity.  However, any discussion of polarisation must be
prefaced with the caveat that there will be considerable uncertainty in the
emitted polarisation, \eg, due to the emission process (synchrotron or Compton
scattering), and the geometry of the emitting region (tangled magnetic
fields, thick accretion disks, \etc).  Because of this large uncertainty,
a simple fiducial polarisation model is adopted.  For the purpose of
investigating the generic effects of strong lensing upon the polarisation,
the emitted polarisation fraction is set to be constant and polarised
orthogonally to the spin axis of the hole (also the orbital axis).  Such a
geometry might
be expected, \eg, when the primary emission mechanism is synchrotron, and
the magnetic field is vertically aligned.  The polarisation may be
substantially reduced in the presence of realistic field geometries,
alternate polarising mechanisms, or radiative transfer effects such as
Faraday rotation or depolarisation.  Nonetheless, this provides some
insight into the possible complexity that may arise in the polarised
spectrum.

\section{Light Curves} \label{LC}
Although this paper is primarily concerned with resolved images, considerable
information can be extracted from the unresolved unpolarised and polarised
light curves.  Since obtaining light curves doees not require imaging capabilities,
and thus is likely to be technically easier, they are discussed first.

\subsection{Unpolarised Flux}
In general, the shape of the light curve in its entirety is required to
make statements regarding the parameters of the black hole and the orbit.
Nevertheless, some of these parameters leave their strongest signatures on
particular portions of the light curves.

\begin{figure}
\begin{center}
\includegraphics[width=\columnwidth]{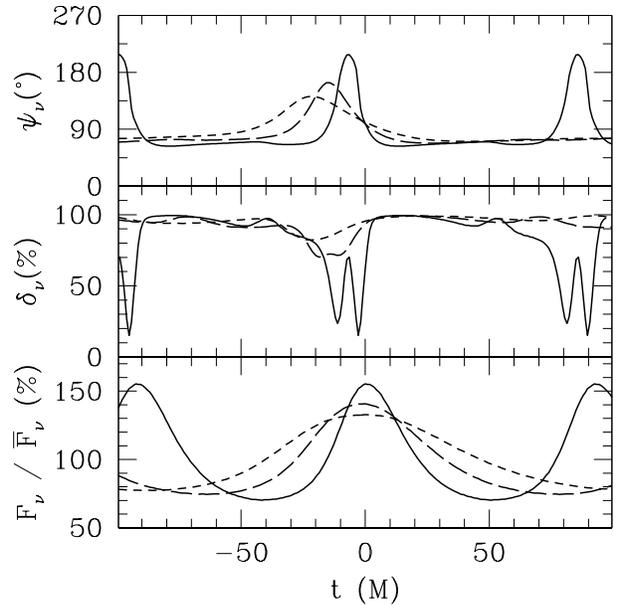} 
\end{center}
\caption{The magnification (bottom), polarisation fraction (middle), and
polarisation angle (top) as functions of time for orbital radii of $6M$
(solid), $8M$ (long dash), and $10M$ (short dash) around a Schwarzschild
black hole viewed from $45^\circ$ above the orbital plane.  A
polarisation angle of $180^\circ$ is orthogonal to the orbital plane
(see Figure \ref{pol_maps}).
The time axis is set so that a single orbital period of the $10M$ case
is shown.  For a black hole mass of $4\times 10^6{\rm M}_\odot$ (as in
\SgrA), the time unit is $M=20\,\s$.}
\label{rcI}
\end{figure}

Any gravitational lensing transient is characterised by a time scale and a
magnification.  For a compact bright spot on a circular orbit about a
black hole, the former is set by the period of the orbit,
\begin{equation}
P = 2\pi \left( r_S^{3/2} + a \right)\,,
\label{period_eq}
\end{equation}
where $r_S$ is the orbital radius of the spot and a positive Kerr
spin parameter, $a$, corresponds to prograde orbits.  For a given $a$,
it is straightforward to determine the radius of the orbit.  This is
readily apparent in Figure \ref{rcI} in which the magnification (\ie,
the integrated flux normalised by its time-averaged value) is plotted
as a function of time for orbital radii of $6M$, $8M$, and $10M$
around a Schwarzschild black hole.

\begin{figure}
\begin{center}
\includegraphics[width=\columnwidth]{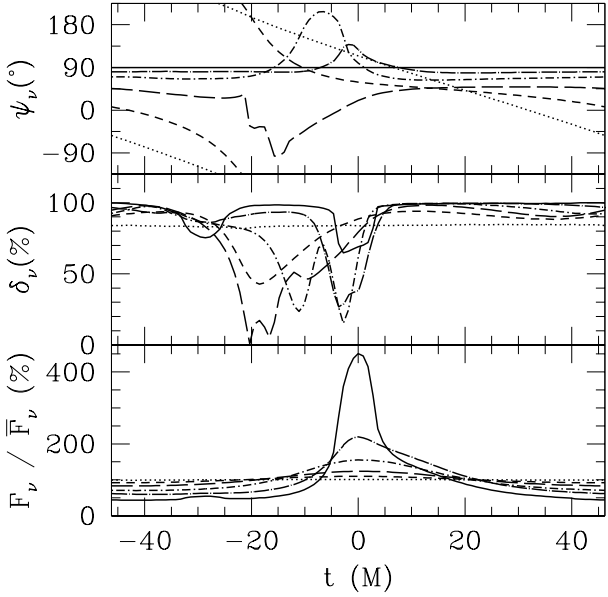} 
\end{center}
\caption{The magnification (bottom), polarisation fraction (middle), and
polarisation angle (top) as a function of time for an orbit at $6M$ around
a Schwarzschild black hole viewed from $0^\circ$ (solid), $22.5^\circ$
(long dash-dot), $45^\circ$ (dash-dot), $67.5^\circ$ (long dash),
$80^\circ$ (short dash) and $89^\circ$ (dotted) above the orbital plane.
The time axis is set so that a single orbital period is shown.}
\label{thcI}
\end{figure}

While the magnification does vary with orbital radius, it is a
stronger indicator of the inclination of the orbit relative to the
line of sight (not to be confused with orbits lying out of the
equatorial plane of a Kerr black hole).  This is shown in Figure
\ref{thcI} in which the magnification is shown for viewing angles
ranging from edge on ($0^\circ$) to nearly face on ($89^\circ$).  In
addition to the strong magnification for orbits which pass directly
behind the black hole, there is a second feature near $t\simeq-30M$
resulting from those null geodesics which make a complete orbit before
escaping to infinity.  The significance of these in the context of the
Fe K$\alpha$ lines has been previously noted by \citet{Beck-Done:05}.
Here it provides a second signature of the strong bending of light,
though only for orbits that are viewed nearly edge on.

\begin{figure}
\begin{center}
\includegraphics[width=\columnwidth]{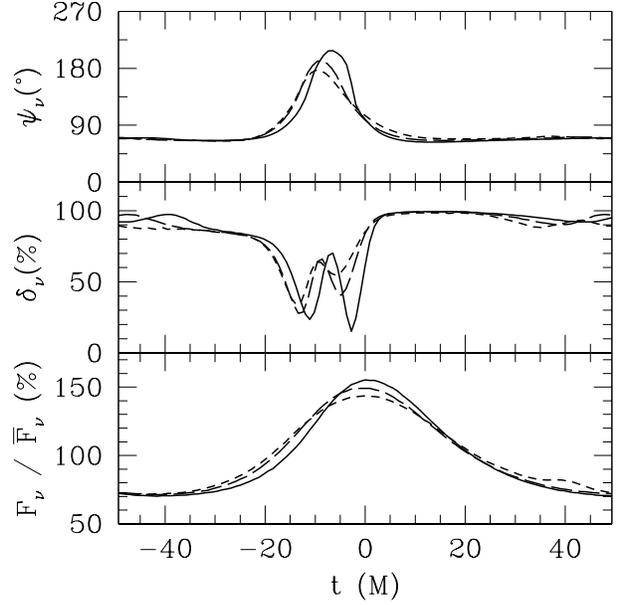} 
\end{center}
\caption{The magnification (bottom), polarisation fraction (middle), and
polarisation angle (top) as a function of time for an orbit at $6M$ around
a Kerr black hole with $a=0$ (solid), $0.5$ (long dash), and $0.998$ (short
dash) viewed from $45^\circ$ above the orbital plane.  The time axis is set
so that a single orbital period of the $a=0.998$ case is shown.}
\label{acI}
\end{figure}

\begin{figure}
\begin{center}
\includegraphics[width=\columnwidth]{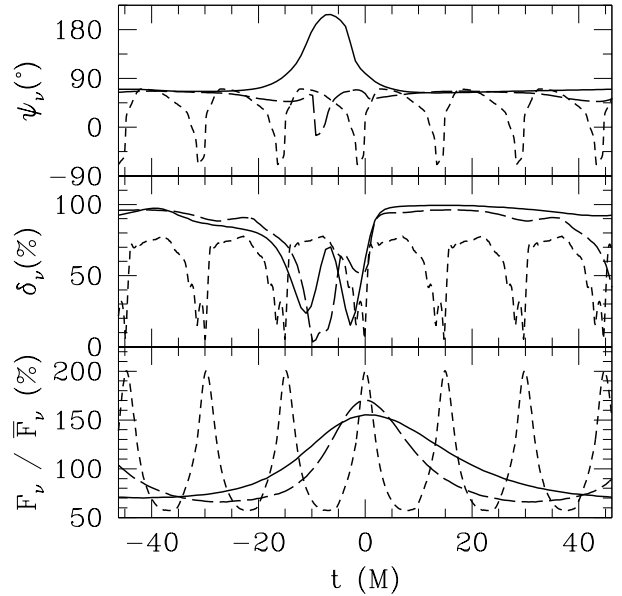} 
\end{center}
\caption{The magnification (bottom), polarisation fraction (middle),
  and polarisation angle (top) as a function of time for orbits at the
  prograde ISCO of the Kerr space-time $a=0$ (solid), $0.5$ (long dash),
  and $0.9$ (short dash) viewed from $45^\circ$ above the orbital
  plane.  The time axis is set so that a single orbital period of the
  $a=0$ case is shown.}
\label{piscocI}
\end{figure}

The dependence of the magnification light curve upon $a$ is presented in
Figures \ref{acI} and \ref{piscocI}.  In Figure \ref{acI}, different spin
parameters ($a=0$, $0.5$, $0.998$) are compared at a fixed radius in
Boyer-Lindquist coordinates, namely $6M$.  In this case, higher black hole
spins tend to demagnify the source slightly.  However, this may be of
secondary significance when compared to the fact that for higher $a$,
stable circular orbits extend closer to the horizon.  Indeed, as
seen in Figure \ref{piscocI}, placing the orbits at the ISCO
results in a net increase in the maximum magnification (and a substantially
reduced orbital period!).

\begin{figure}
\begin{center}
\includegraphics[width=\columnwidth]{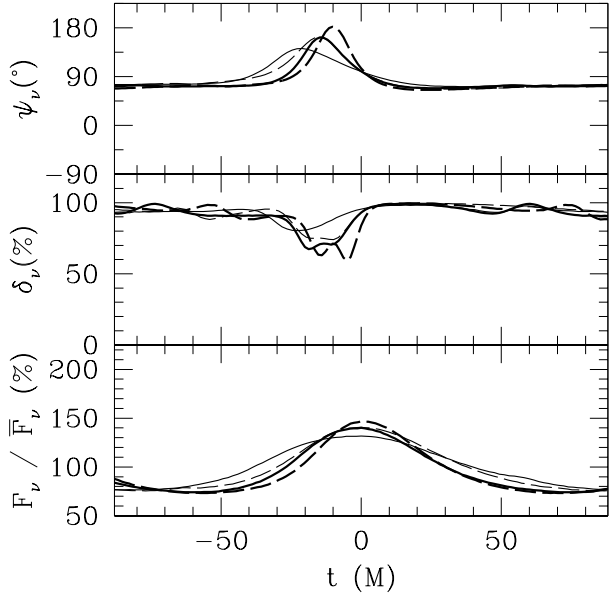} 
\end{center}
\caption{The magnification (bottom), polarisation fraction (middle), and
polarisation angle (top) as a function of time for prograde (solid line)
and retrograde (long-dash line) orbits at the retrograde ISCO of the Kerr
space-time for $a=0.5$ (thick line) and $a=0.998$ (thin line) viewed from
$45^\circ$ above the orbital plane.  The time axis is set so that a single
orbital period of the prograde $a=0.998$ case is shown.  In all cases the
direction of the orbital motion is the same (\ie, the retrograde orbits are
generated by reversing the spin of the hole).}
\label{riscocI}
\end{figure}

For rapidly rotating black holes one would expect a difference in the
lensing signature of prograde and retrograde orbits.  This will be
mitigated somewhat by the fact that the retrograde ISCO moves out
considerably, reaching $\sim 9M$ for the maximally rotating case.
Nonetheless, as seen in Figure \ref{riscocI}, there are quantitative
differences in the magnification similar to those seen when the spin is
varied (\cf~Figure \ref{acI}) as expected.

\subsection{Polarised Flux}
In addition to the magnification, Figures \ref{rcI}--\ref{piscocI} also
show the fractional change in the emitted polarisation and the polarisation
angle for the fiducial polarisation model described in section \ref{CM:OTS}
(and subject to the caveats presented at the end of that section).  In
general, the polarisation fraction and polarisation-angle light curves show
considerable structure.  The variability results primarily from special
relativistic aberration associated with the rapid orbital motion, coupled
with the choice of the emitted polarisation.  This effect is substantially
amplified by both gravitational lensing (which allows many viewing angles
to be sampled) and general relativistic transfer effects (which rotate the
polarisation direction according to the rules of parallel propagation).

The two primary features in the polarisation light curves are the decrease
in polarisation fraction (referred to in what follows as the primary
minimum) and rotation of the polarisation angle immediately preceding the
maximum magnification.  Both phenomena result from the development
of an Einstein ring/arc, and are thus expected to be generic features.  For the
configuration considered here, these are a strong function of radius (see,
\eg, Figure \ref{rcI}) and viewing angle (Figure \ref{thcI}).  The
dependence upon viewing angle reaches a maximum when a substantial portion of
the rays are incident along the orbital axis (note that due to
gravitational lensing this does not occur when the orbit is face on).

A third notable feature is the location of a second minimum due to the
development of a secondary ring/arc resulting from rays which complete an orbit
before escaping to infinity.  This second minimum is considerably shallower
than (and lags substantially behind) the primary minimum and should not be
confused with the substructure within the primary minimum.  The decrease in
the polarisation fraction results for reasons similar to those that
produce the primary minimum, and is similarly expected to be a generic
feature.  It is present in Figures \ref{rcI} and \ref{thcI}, where it leads
the primary minimum by $\sim 50M$.  However, of greater interest is the
strong dependence of this feature upon the structure of the space-time near
the photon orbit, and thus its sensitivity to strong-field
general-relativity.  This is explicitly demonstrated in Figures \ref{acI}
and \ref{riscocI}, in which the lag between the primary and secondary
minima is a strong function of black hole spin.

As the orbit becomes more relativistic, a permanent partial ring
develops, qualitatively changing the polarisation features.  This can
be seen in Figure \ref{piscocI}.  However, it can be more clearly seen
in the images, [\eg, Figure \ref{image_seq}, panel (e)], and thus is
discussed in the following section.

\section{Images} \label{Im}
In addition to measuring the light curves, future sub-millimetre
observations promise to image the inner regions of the Galactic centre at
$20\,\muas$ resolution \citep{Gree:05,Miyo_etal:04}.  Illustrative
images of the orbits are presented in this section.  These may be
compared with calculations which assume a uniform and steady accretion
flow \citep[\eg,][]{Falc-Meli-Agol:00,Taka:04,Taka:05}.

\begin{figure*}
\begin{center}
\begin{tabular}{cc}
\includegraphics[width=0.5\textwidth]{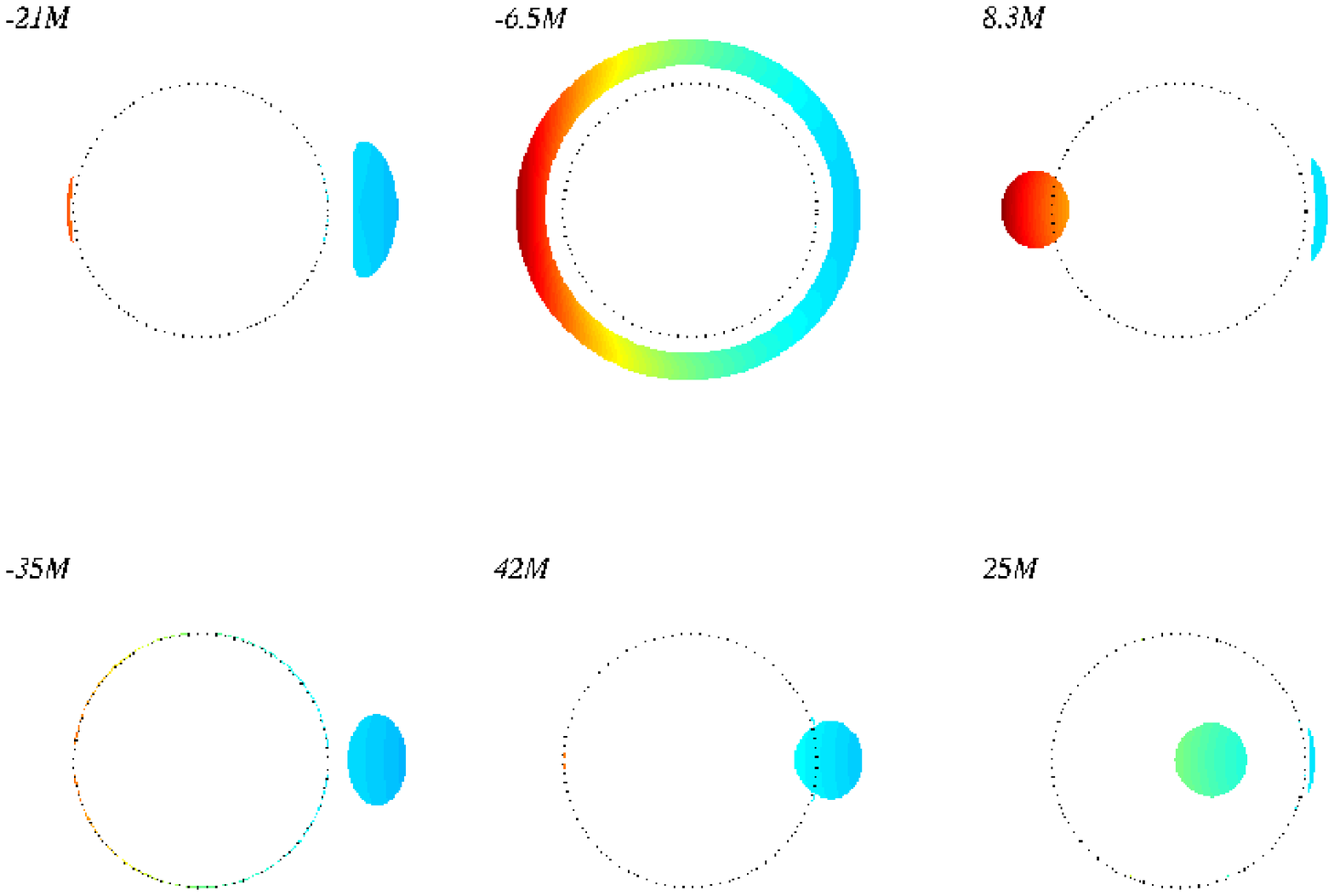} 
&
\includegraphics[width=0.5\textwidth]{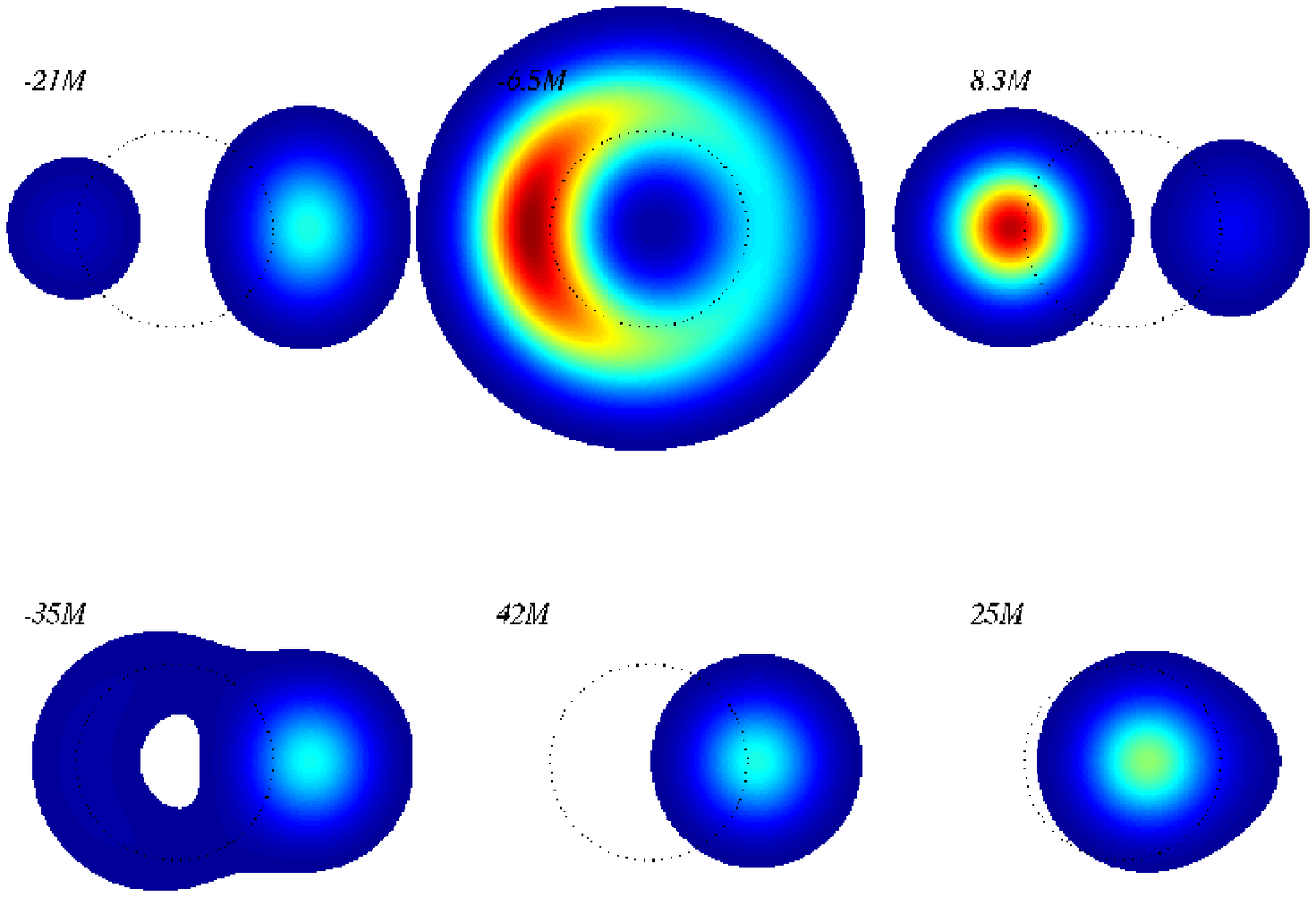} 
\\
(a) & (b)\\
&\\
&\\
\includegraphics[width=0.5\textwidth]{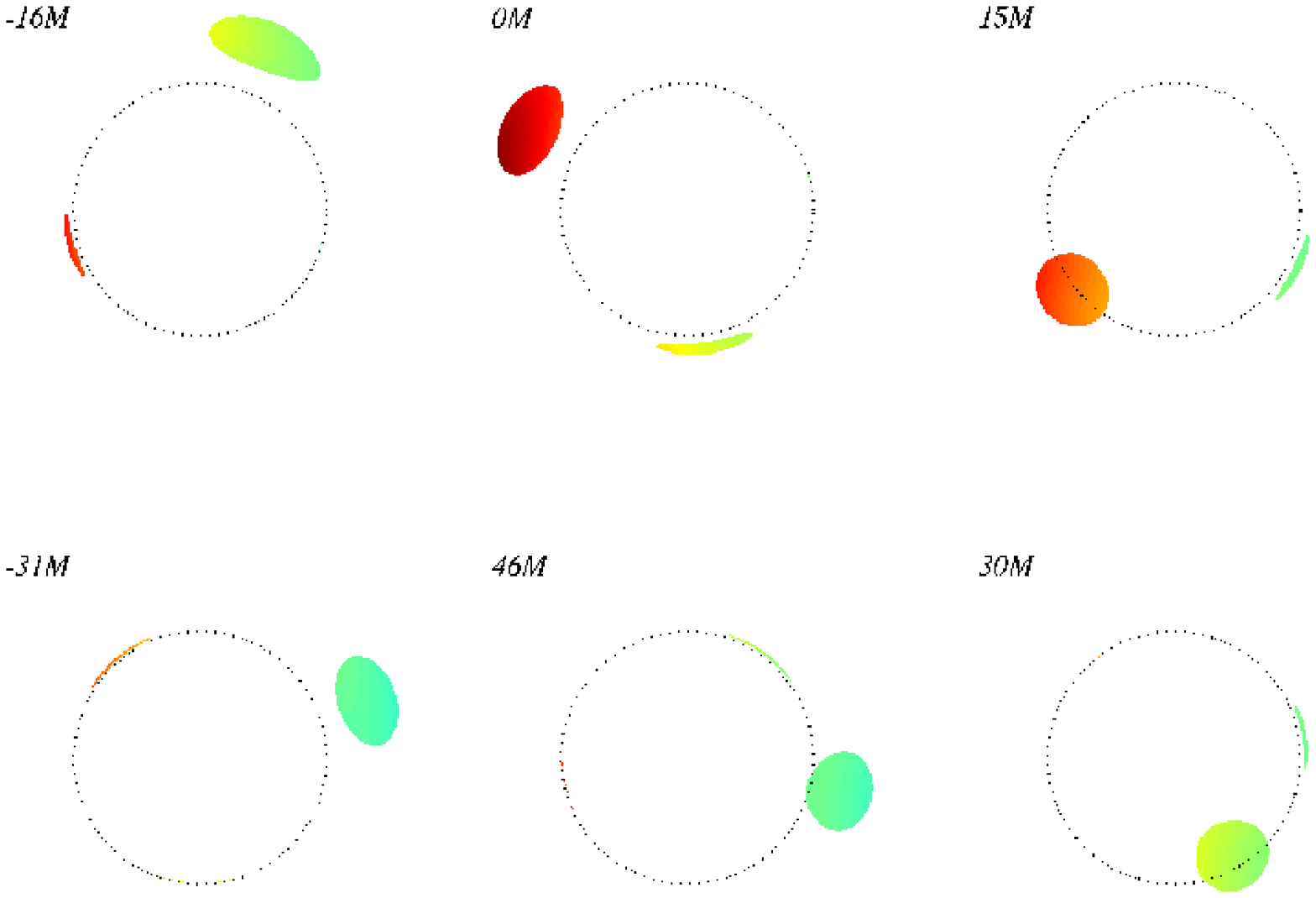} 
&
\includegraphics[width=0.5\textwidth]{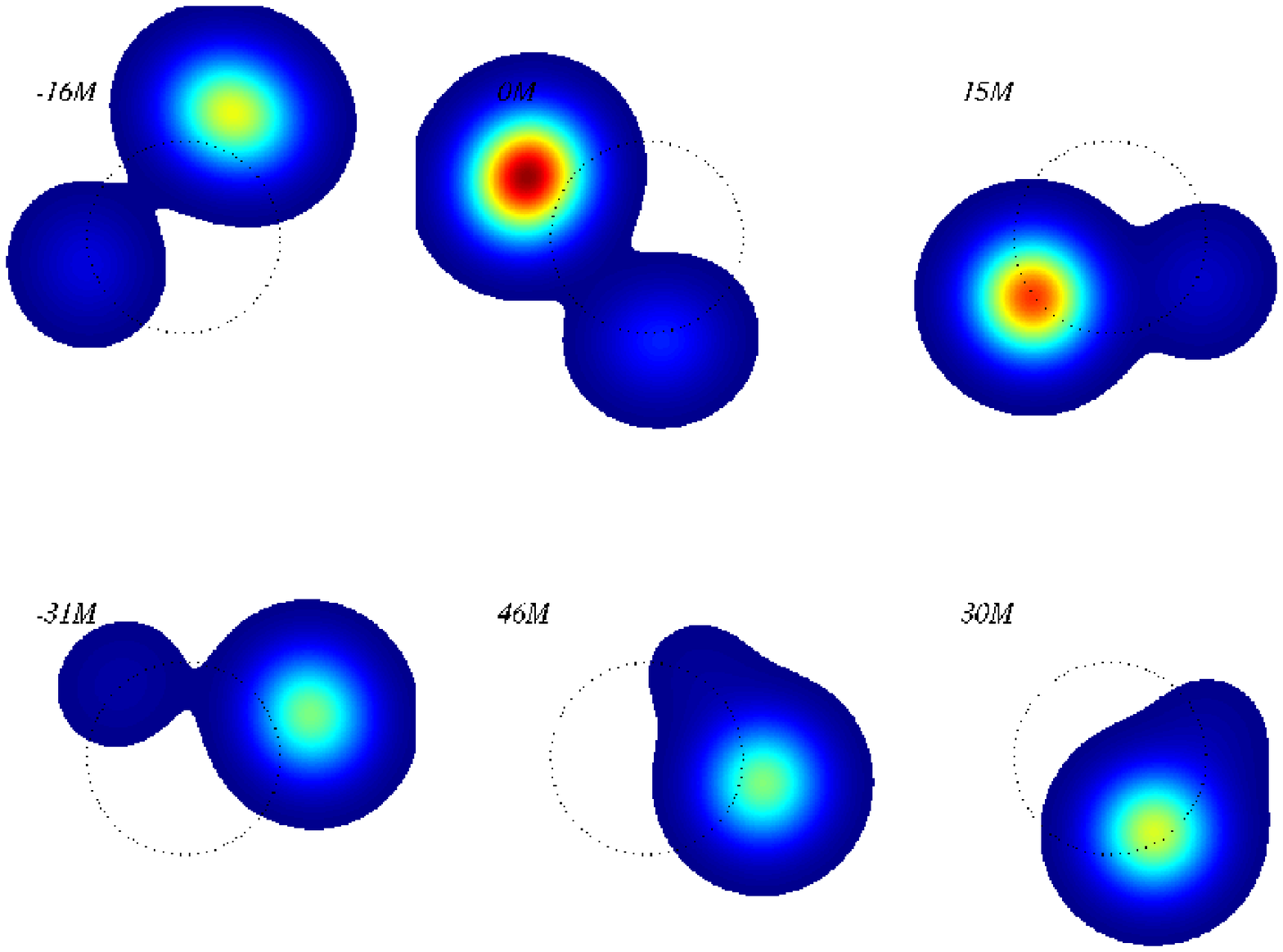} 
\\
(c) & (d)\\
&\\
&\\
\includegraphics[width=0.5\textwidth]{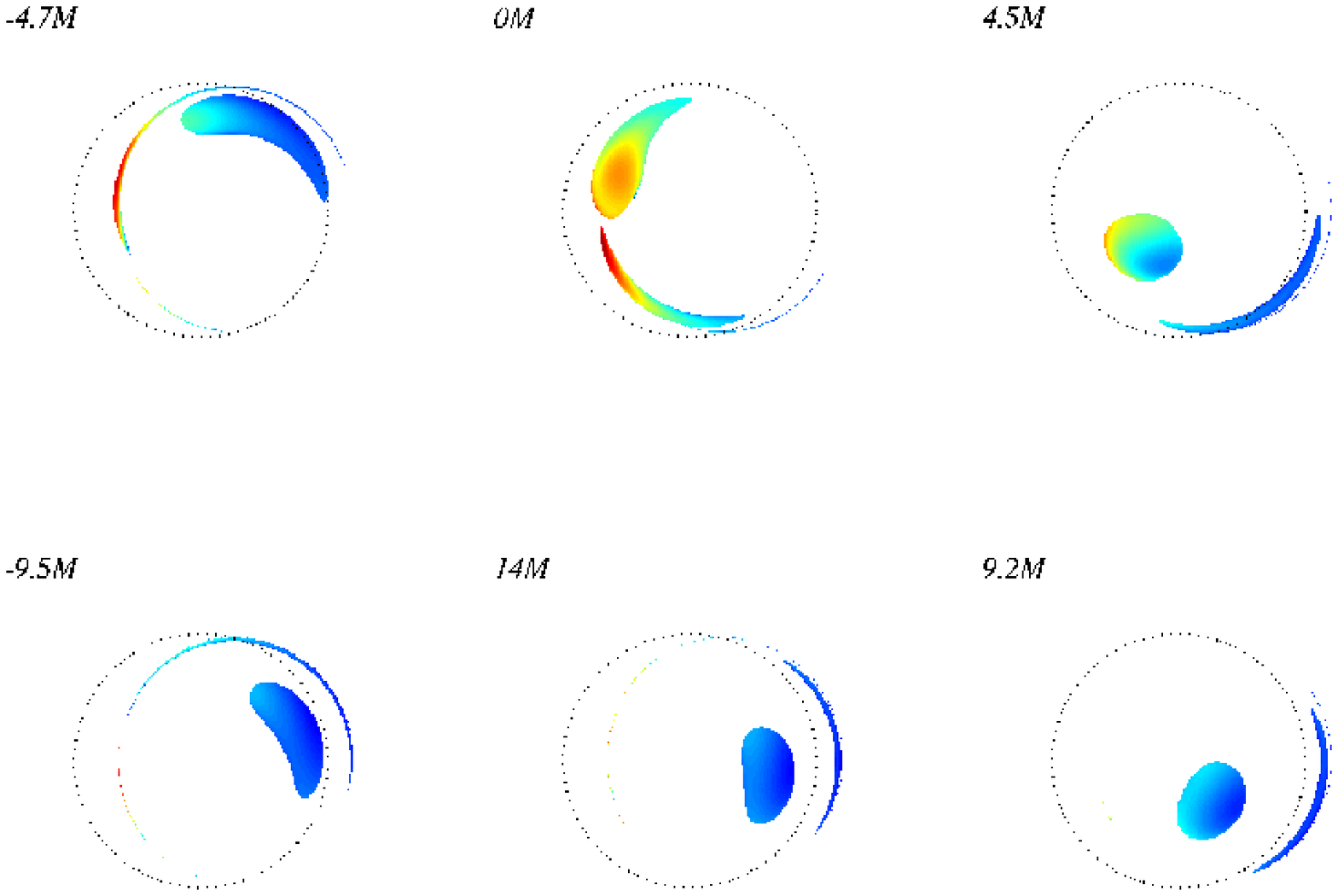} 
&
\includegraphics[width=0.5\textwidth]{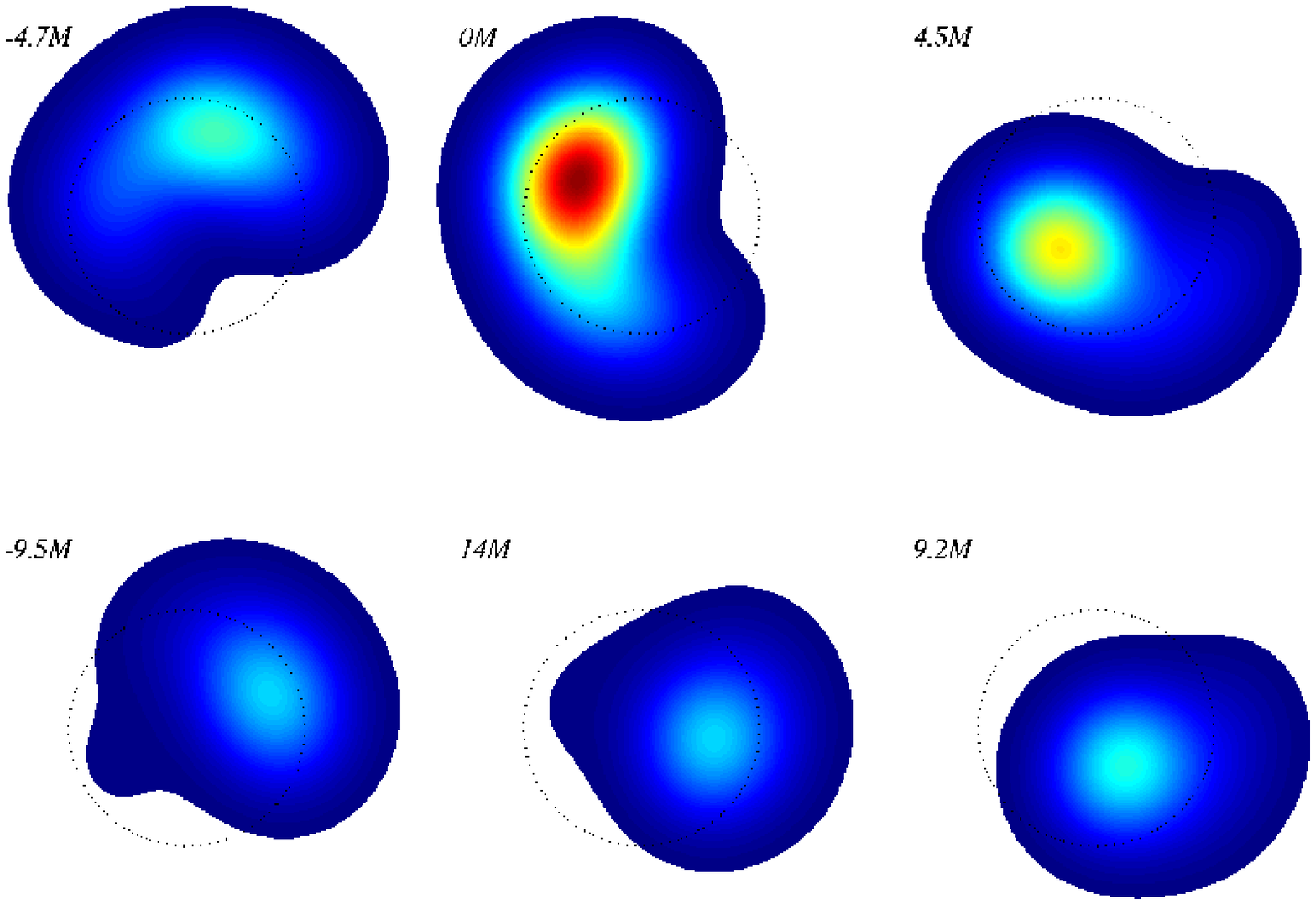} 
\\
(e) & (f)\\
\end{tabular}
\end{center}
\caption{Illustrative sequences of intensity maps.  Cases (a) \& (b) refer
to $r_S=6M$, $a=0$, viewed in the orbital plane, (c) \& (d) refer to
$r_S=6M$, $a=0$, viewed from $45^\circ$ above the orbital plane, and (e) \&
(f) refer to an orbit at the prograde ISCO for $a=0.9$, also view from
$45^\circ$ above the orbital plane.  Panels on the left are shown at the
computed resolution, while panels on the right are smoothed by an
instrument resolution of $20\,\muas$ (achievable with a future sub-mm
VLBA).  For reference, the time as shown in the light curves is listed
in the upper-left corner of each map, and the photon capture impact
parameter ($\sqrt{27}M$) is shown by the dotted circle.}
\label{image_seq}
\end{figure*}

Figure \ref{image_seq} shows a sequence of six images distributed roughly
evenly throughout an orbit for three of the cases discussed in the
previous section.  In particular the upper panels show an orbit around a
Schwarzschild black hole at $r_S=6M$ viewed edge on, the middle panels show
the same orbit viewed from $45^\circ$ above the orbital plane, and the
bottom panels show the prograde ISCO of a Kerr black hole with $a=0.9$
viewed from $45^\circ$ above the orbital plane.  Panels on the left show
the computed resolution, while panels on the right are smoothed by a
Gaussian filter with full width $4M$ (corresponding to $20\muas$ for \SgrA)
and thus are comparable to those that may be observed within the next
decade.  Each sequence of images is separately normalised.

While all three cases appear qualitatively different, they share three
generic features: brightening due to relativistic beaming, a primary
Einstein ring/arc, and a secondary Einstein ring/arc due to photons
which complete at least one orbit around the black hole (also
complete only on the top panel).  The first two features are responsible for the
magnification and primary minimum in the unpolarised and polarised light
curves, respectively.  The third is largely responsible for the secondary
minimum in the polarisation fraction.  Since the secondary ring/arc is formed
by photons which pass near the photon orbit ($3M$ for Schwarzschild, $M$
for maximally rotating Kerr; but see the footnote in section \ref{I}),
it provides a sensitive test of strong field relativity.  As a direct
result, the position of the secondary minimum is a diagnostic of the
black hole spin, as found in section \ref{LC}.

Despite the commonalities, there are considerable differences between the
three cases shown.  The most obvious is that the primary and secondary
Einstein rings are incomplete for orbits viewed significantly out of the
orbital plane.  This results in reduced magnification and polarisation
deviations in the lightcurves.  In the high spin case, a permanent
incomplete ring develops resulting in a qualitative change in the polarised
light curves.

As seen in the right-hand panels, these features remain in the smoothed
images.  Therefore, imaging may be a diagnostic of the orbital and black
hole parameters.

\begin{figure*}
\begin{center}
\begin{tabular}{ccc}
\includegraphics[width=0.31\textwidth]{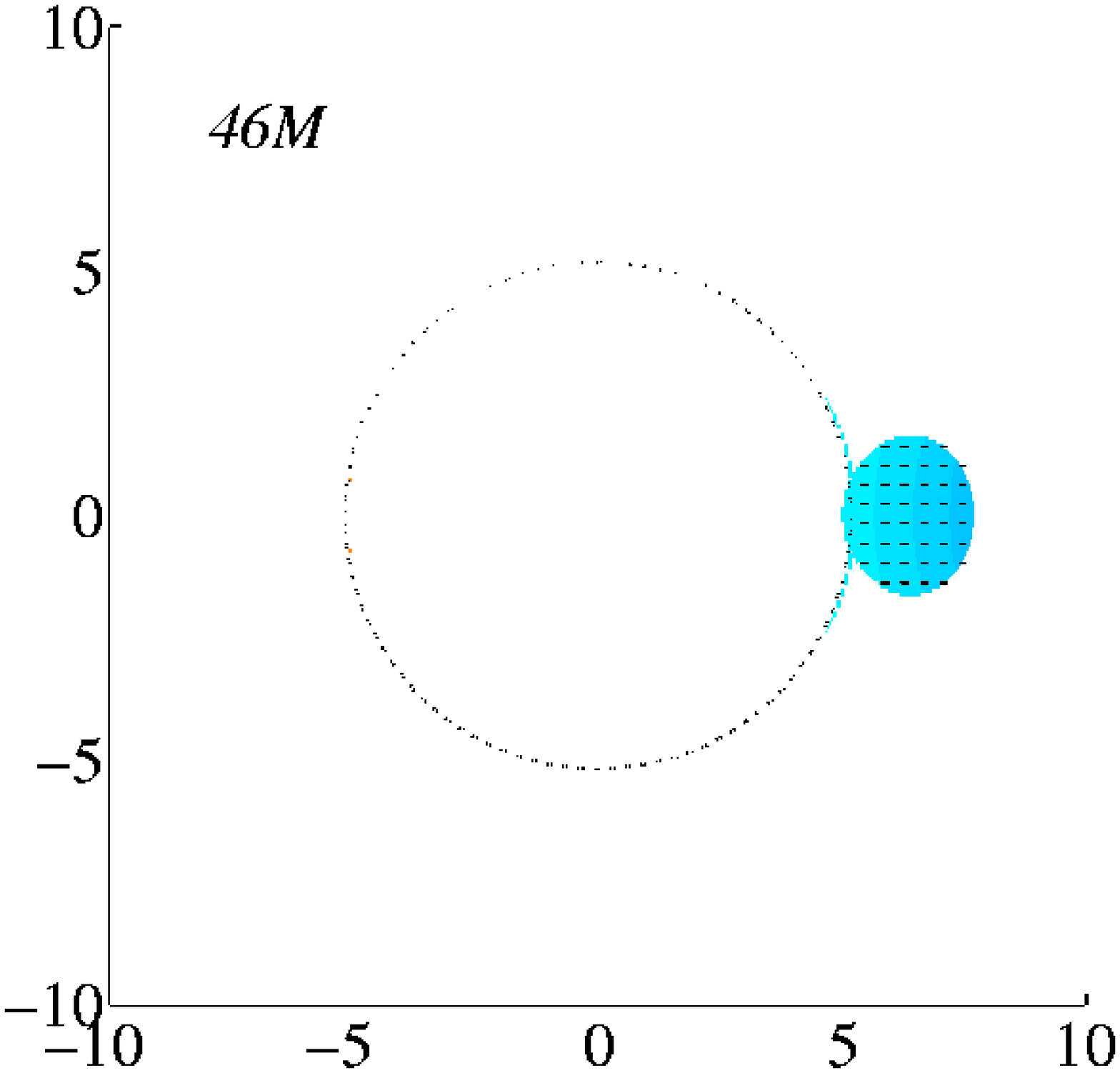} 
&
\includegraphics[width=0.31\textwidth]{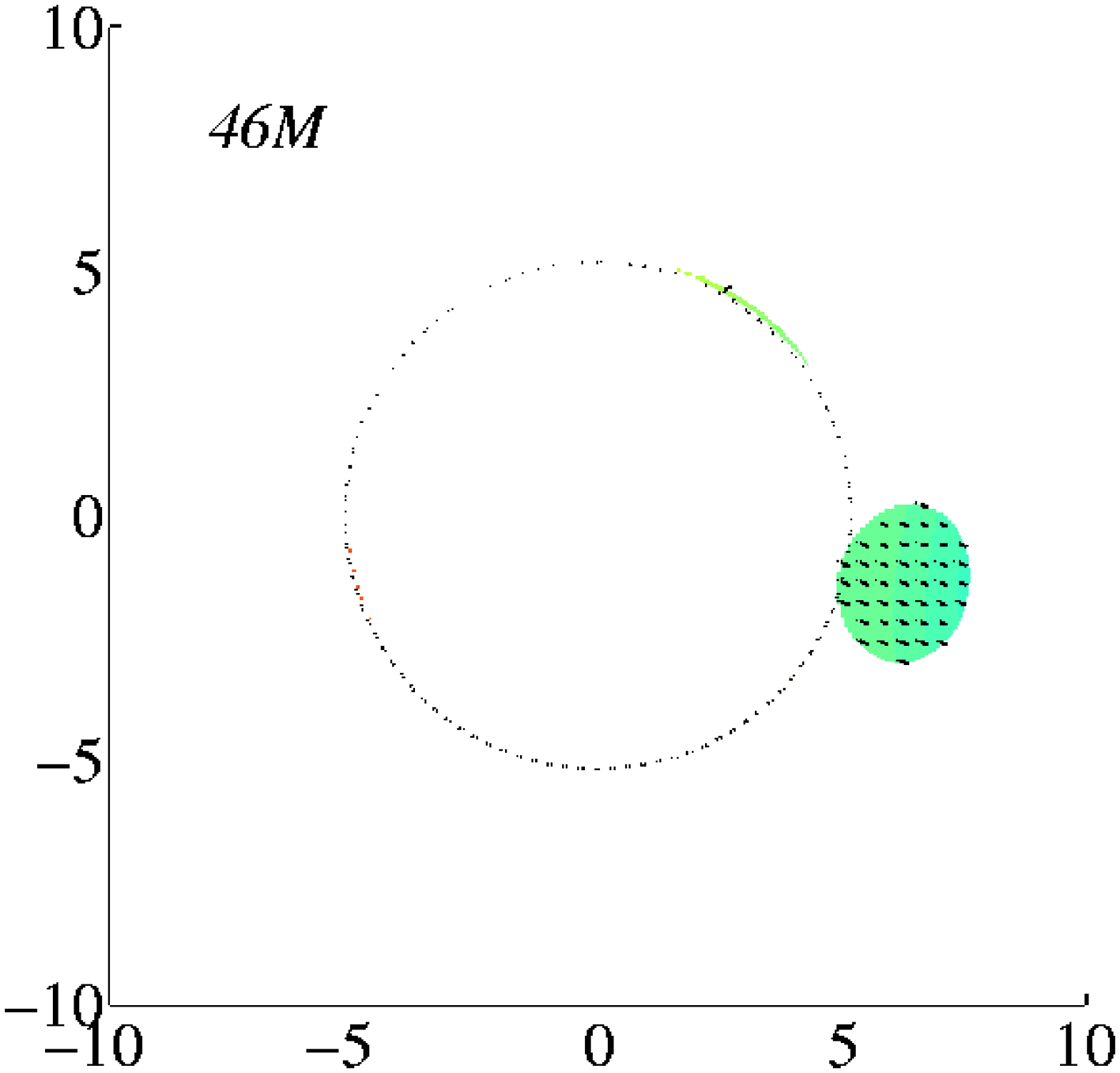} 
&
\includegraphics[width=0.31\textwidth]{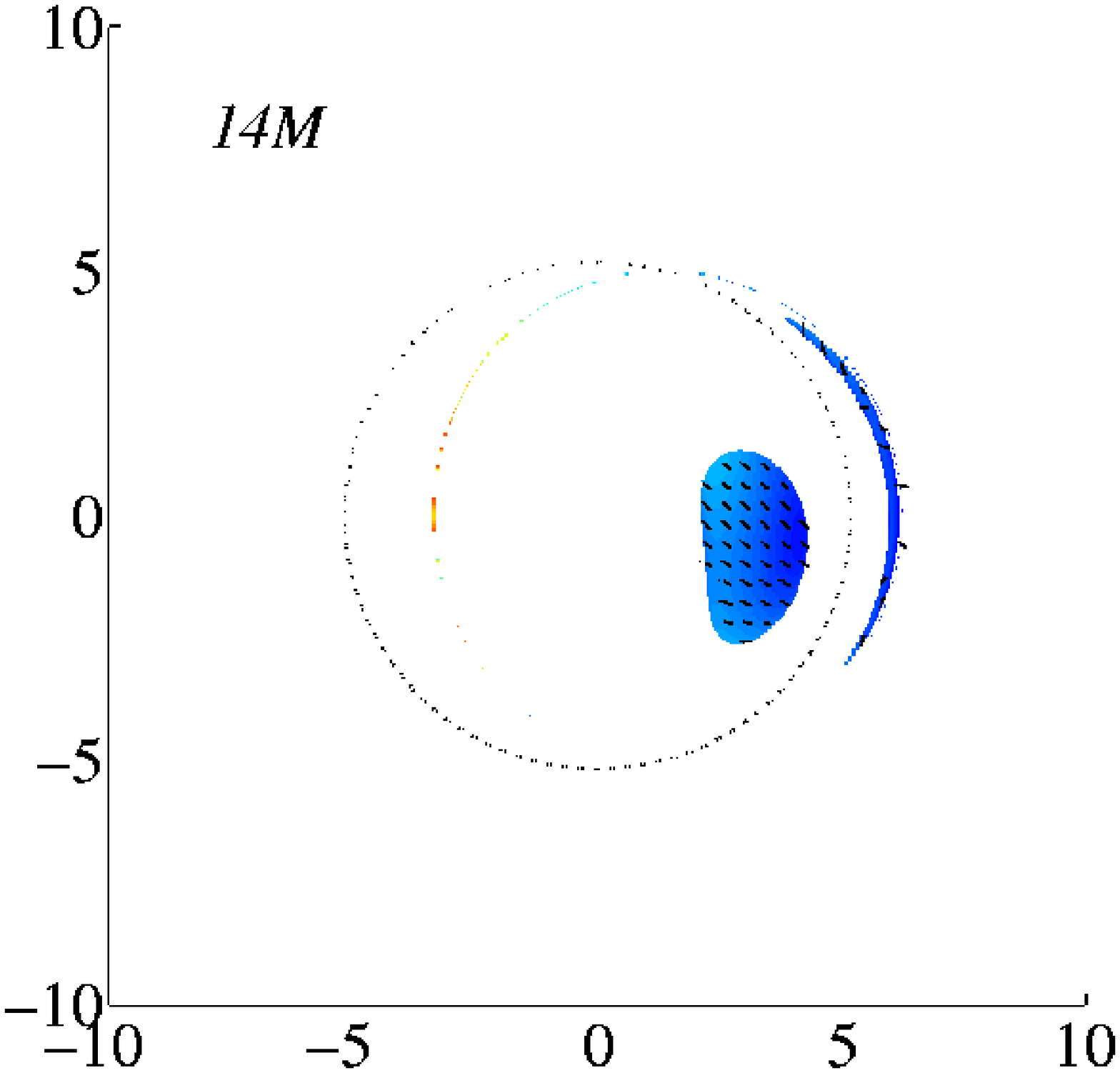} 
\\
(a) & (b) & (c)\\
&&\\
&&\\
\includegraphics[width=0.31\textwidth]{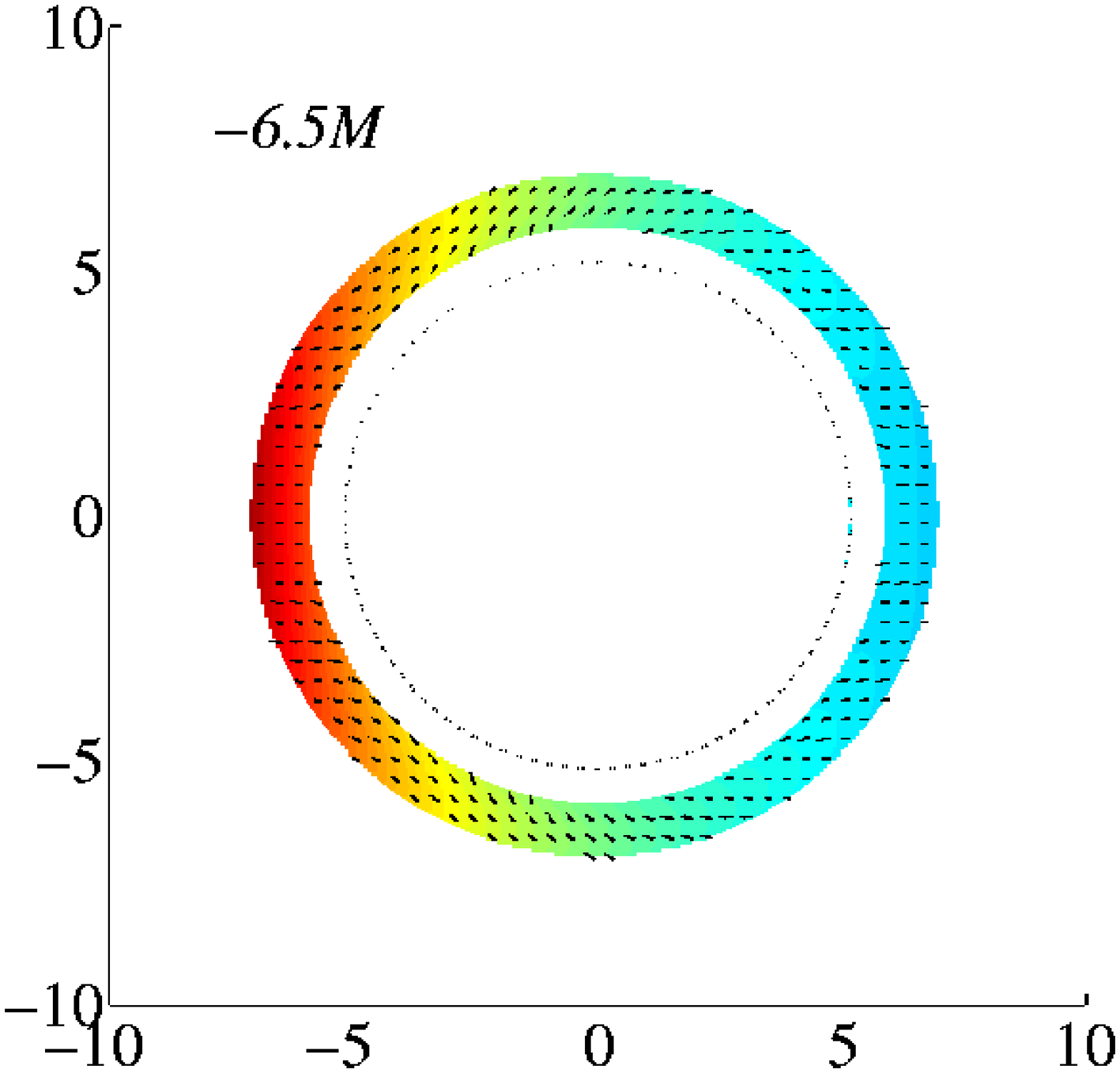} 
&
\includegraphics[width=0.31\textwidth]{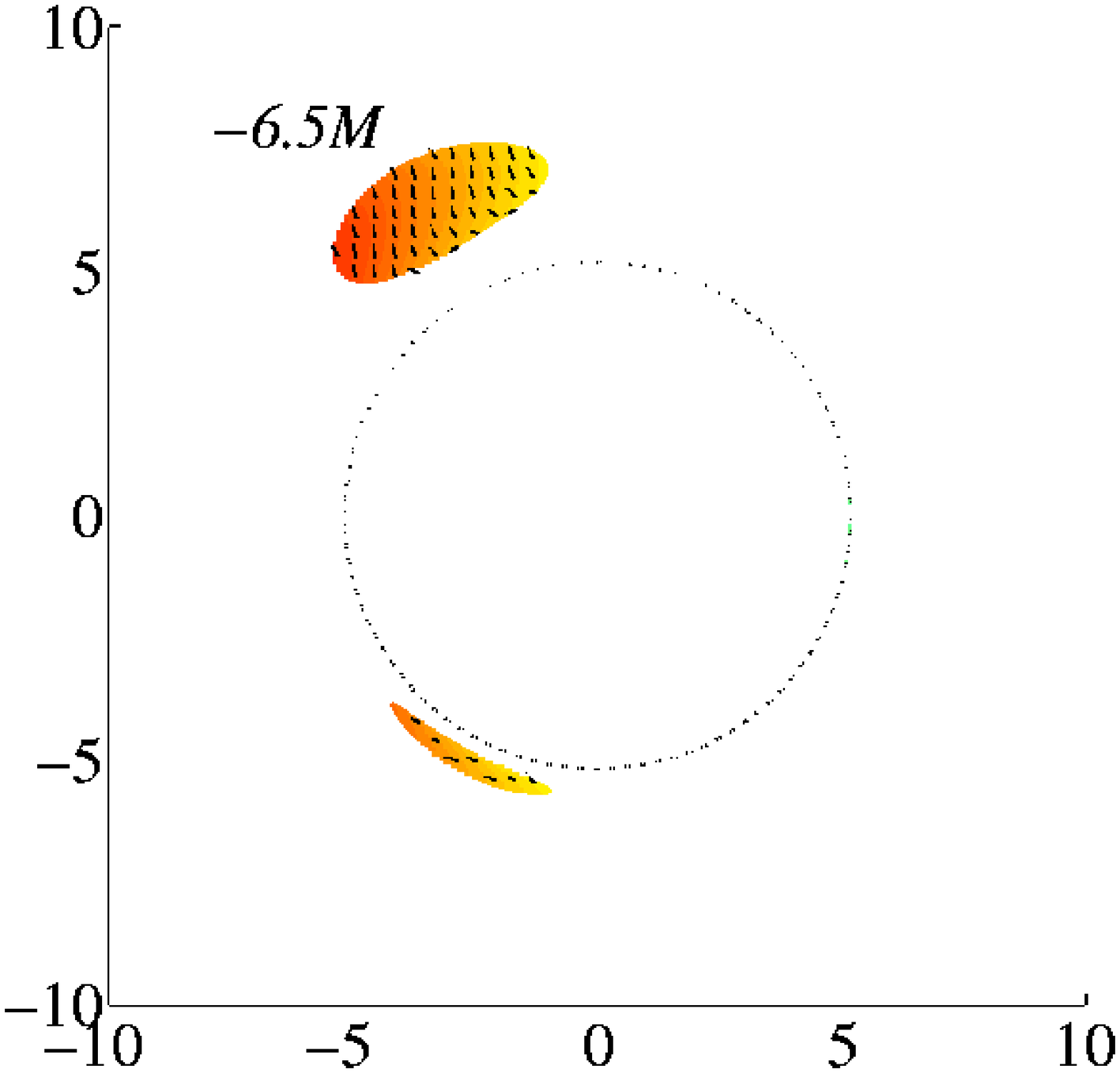} 
&
\includegraphics[width=0.31\textwidth]{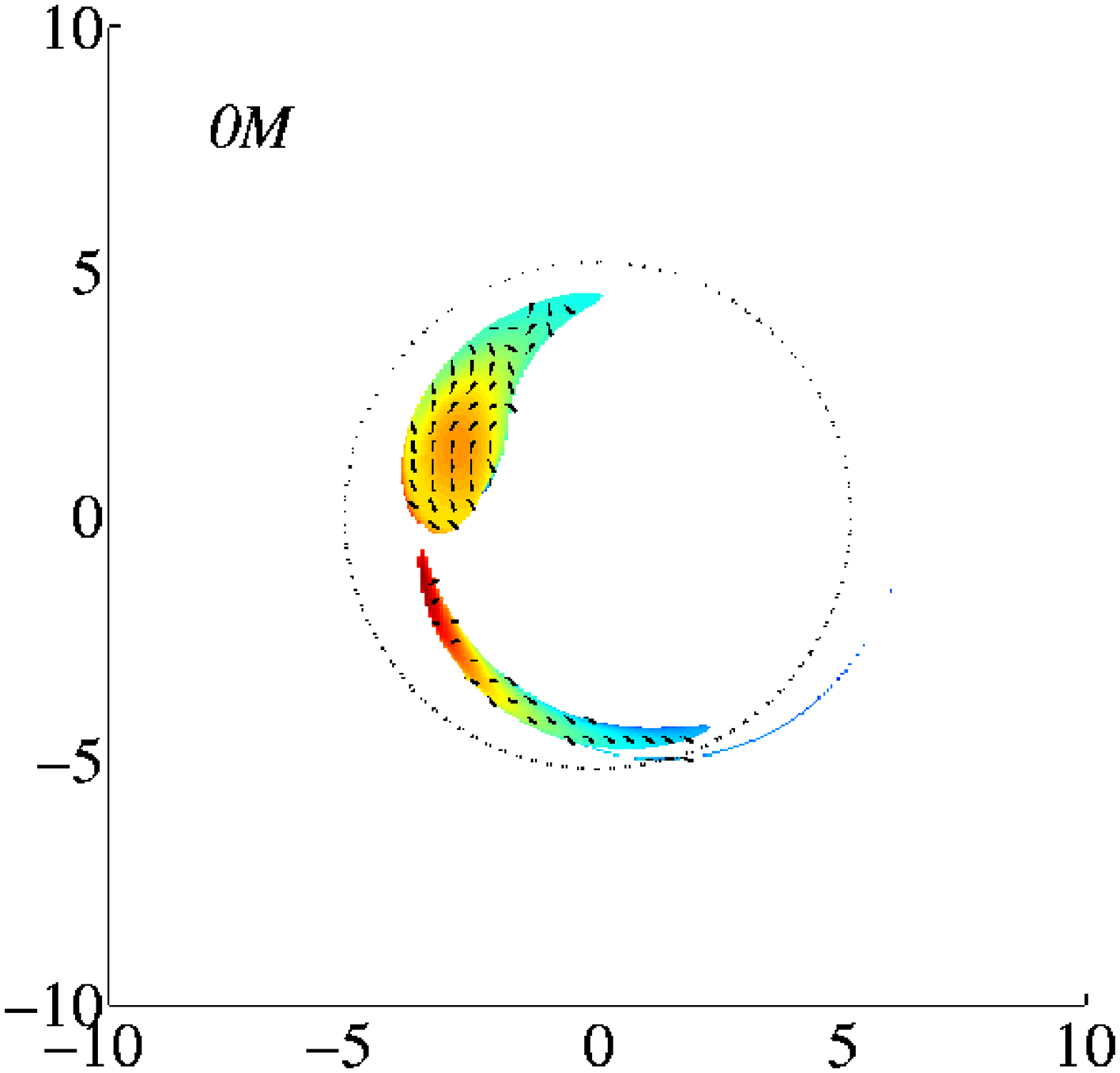} 
\\
(d) & (e) & (f)
\end{tabular}
\end{center}
\caption{Illustrative polarisation maps for the cases shown in Figure
\ref{image_seq}.  Cases (a) \& (d) refer to $r_S=6M$, $a=0$, viewed in the
orbital plane, (b) \& (e) refer to $r_S=6M$, $a=0$, viewed from $45^\circ$
above the orbital plane, and (c) \& (f) refer to an orbit at the prograde
ISCO for $a=0.9$, also viewed from $45^\circ$ above the orbital plane.  The
top panels show the polarisation when the emitting sphere is $180^\circ$ out
of phase with the maximum magnification and thus are indicative of the
unlensed polarisation.  The bottom plots are illustrative of the
polarisation structures within the primary minima in polarisation angle.
For reference, the photon capture impact parameter ($\sqrt{27}M$) is shown by
the dotted circle.  Axes are labelled in units of $M$.  For a mass of
$4\times10^6{\rm M}_\odot$ and distance $8\,\kpc$, appropriate for \SgrA,
the scale $M$ corresponds to $5\,\muas$.}
\label{pol_maps}
\end{figure*}

The use of the polarisation as a diagnostic of the space-time structure is
illustrated in Figure \ref{pol_maps}, in which images of the
polarisation angles are overlayed upon the intensity maps.  For the
purpose of comparison, these are shown when the emitting sphere is
$180^\circ$ out of phase with the maximum magnification (and thus
typical of the unlensed polarisation, top) and for a typical point
within the primary polarisation minimum (bottom).  As seen in the
bottom panels, the development of a primary Einstein ring/arc
generically rotates the polarisation angle and reduces the integrated
polarisation flux.  The development of a secondary ring/arc yields
similar results, and hence polarisation is sensitive to those image
features which probe strong gravity most effectively.  Therefore,
polarisation can be use to probe the spacetime at the photon orbit, as
long as its emission and transfer are understood.

\begin{figure}
\begin{center}
\includegraphics[width=\columnwidth]{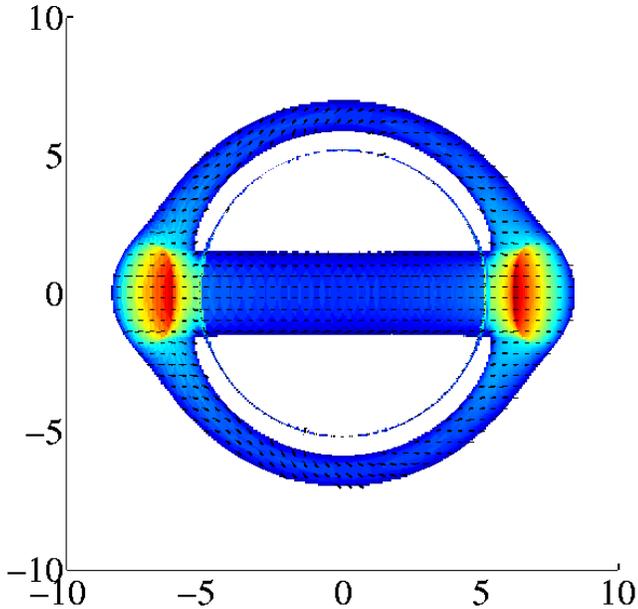} 
\end{center}
\caption{The time average of the intensity and polarisation maps for an
edge-on orbit of an emitting sphere at $6M$ around a Schwarzschild black
hole. For reference, the photon capture impact parameter ($\sqrt{27}M$) is
shown by the dotted circle.  Axes are labelled in units of $M$.}
\label{p6_a0_th90_avg}
\end{figure}

\begin{figure}
\begin{center}
\includegraphics[width=\columnwidth]{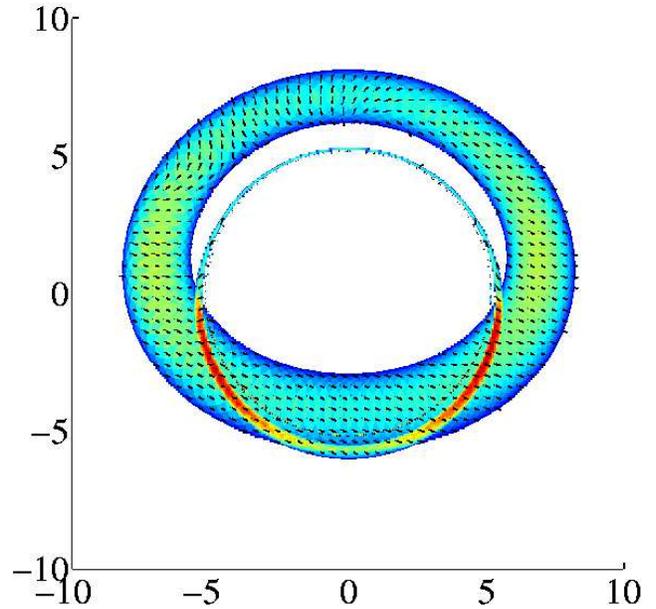} 
\end{center}
\caption{The average of the intensity and polarisation maps for an orbit at
$6M$ around a Schwarzschild black hole viewed from $45^\circ$ above the
orbital plane.  For reference, the photon capture impact parameter
($\sqrt{27}M$) is shown by the dotted circle.  Axes are labelled in units of
$M$.}
\label{p6_a0_th45_avg}
\end{figure}

\begin{figure}
\begin{center}
\includegraphics[width=\columnwidth]{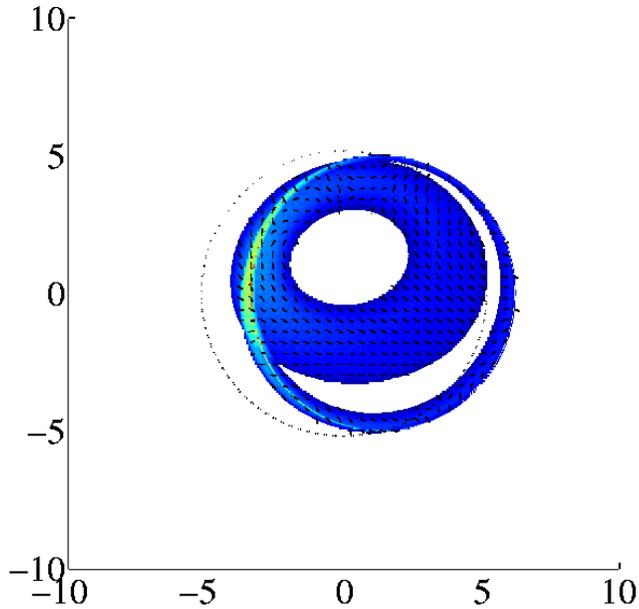} 
\end{center}
\caption{Average of the intensity and polarisation maps for an orbit at the
prograde ISCO around a Kerr black hole with $a=0.9$, viewed from $45^\circ$
above the orbital plane.  For reference, the photon capture impact parameter
($\sqrt{27}M$) is shown by the dotted circle.  Axes are labelled in units of
$M$.}
\label{pISCO_a09_th45_avg}
\end{figure}

\begin{figure}
\begin{center}
\includegraphics[width=\columnwidth]{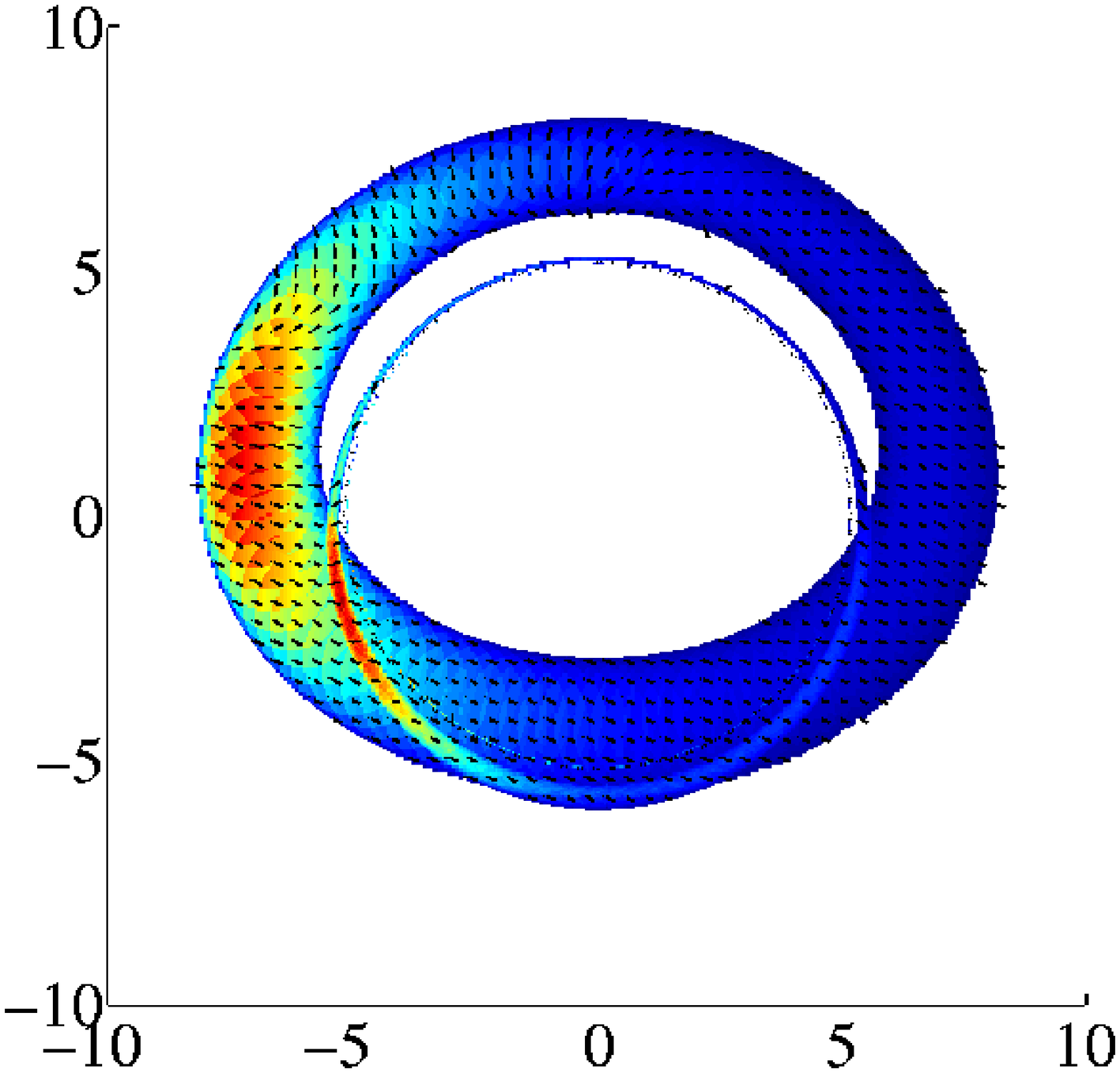} 
\end{center}
\caption{Same as Figure \ref{p6_a0_th45_avg}, with
  $I_\nu\propto\nu^{-1.3}$.  For reference, the photon
  capture impact parameter ($\sqrt{27}M$) is shown by the dotted
  circle.  Axes are labelled in units of $M$.}
\label{p6_a0_th45_avg_synch}
\end{figure}

The images in Figures \ref{image_seq} and \ref{pol_maps} are instantaneous
snapshots, and are only observable through exposures that are considerably
shorter than the orbital period. Typically, observations will average these
images over the instrument integration time. To reflect more realistic
circumstances, Figures \ref{p6_a0_th90_avg}-\ref{pISCO_a09_th45_avg} show
the intensity and polarisation maps averaged over a full orbit (for the
same orbits shown in Figures \ref{image_seq} and \ref{pol_maps}).  These
are produced by summing the images used to compute the light curves in
section \ref{LC}.  All of these have similar structures: a ring/arc associated
with the direct image of the sphere, and a second ring/arc associated with the
secondary Einstein ring/arc (which closely follows the photon capture impact
parameter).

The remarkable symmetry is an artifact of the emission
scheme used; for thermal emission in the Rayleigh-Jeans limit, the
special relativistic red-shift is precisely offset by the apparent
motion on the sky\footnote{This can be explicitly seen in the
  images. Indeed the left side of the images have a lower resolution,
  resulting from the poorer quality of the averages than the right
  side.  This is a direct result of the emitting sphere spending less
  time on the left (where it is brighter) than on the right (where it
  is brighter).}.
This can be explicitly demonstrated for a Schwarzschild black hole by
considering a pair of spots moving at a given angle $\theta$ relative
to the line of sight with velocities $\beta$ and $-\beta$ and time
averaged intensities $\langle I^+_\nu \rangle$ and
$\langle I^-_\nu \rangle$, respectively.  For an emitted spectrum with spectral slope
$\alpha$ (\ie, $I_\nu \propto \nu^{-\alpha}$) the time averaged
intensities are related by
\begin{equation}
\frac{\langle I^+_\nu \rangle}{\langle I^-_\nu \rangle}
=
\frac{I^+_\nu}{I^-_\nu} \frac{v_a^-}{v_a^+}
=
\left( \frac{1+\beta\cos\theta}{1-\beta\cos\theta} \right)^{\alpha+2}\,,
\label{asym_bright}
\end{equation}
where $v_a^\pm = \beta \sin\theta/(1\mp\beta\cos\theta)$ is the apparent
velocity on the sky.  Thus, in the Rayleigh-Jeans approximation,
$\langle I^+_\nu \rangle = \langle I^-_\nu \rangle$ and the images are
indeed expected to be symmetric.  However, for $\alpha>-2$, as is the
case for optically thin synchrotron emission, equation
(\ref{asym_bright}) implies a substantial brightness asymmetry. This
can be seen explicitly in Figure \ref{p6_a0_th45_avg_synch}, which is
identical to Figure \ref{p6_a0_th45_avg} except that $\alpha$ was
taken to be $1.3$ \citep[as suggested by infrared and X-ray flare
  observations,][]{Ecka_etal:04}.  However, despite the considerable
asymmetry, the morphology of the image remains unchanged.  Therefore,
measuring the brightness asymmetry provides a method to probe the
spectrum of the bright spot.  Note that in both cases the symmetry is
clearly broken in the polarisation map, and thus polarisation data may be
used to infer the direction of the source motion.

The images do not show a clear black hole shadow.  This is not a
result of approximating an inhomogeneity by an optically thick sphere,
but rather due to emission geometry.  In particular, the black hole is
not everywhere back lit, and hence at positions on the sky where the
primary emitting region lies in front of the black hole no shadow is
present.  Thus, even when the emission is optically thin, it is
expected that the image of inhomogeneous accretion flows will
qualitatively differ from that produced by a quasi-spherical accretion
flow.

It is significant that the average images of the $a=0$ and $a=0.9$ case
differ substantially (\cf~Figures \ref{p6_a0_th45_avg} and
\ref{pISCO_a09_th45_avg}).  In particular, the relative positions of the
orbital and secondary ring are shifted.  Therefore, the spin of the black
hole leads to a relative change in position, which is considerably simpler
to measure through sub-millimetre interferometry.

\section{Centroids}\label{Ce}
It should be possible to measure the image centroid with greater
precision than the resolution of an image.  As seen in Figure
\ref{image_seq}, the motion of the centroid contains a considerable amount
of the information in the image, and thus the path of the centroid provides a
diagnostic of the orbital and black hole parameters.

\begin{figure}
\begin{center}
\includegraphics[width=\columnwidth]{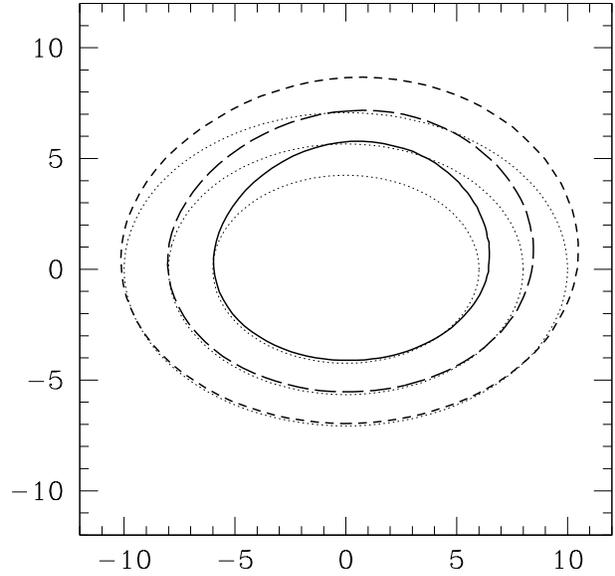} 
\end{center}
\caption{Motion of the centroid of the intensity for circular orbits around
a Schwarzschild black hole viewed from $45^\circ$ above the orbital plane
with radii of $6M$ (solid), $8M$ (long dash), and $10M$ (short dash).  For
reference, a circle inclined at $45^\circ$ is also shown by the dotted
lines.  Axes are labelled in units of $M$.}
\label{r_cents}
\end{figure}

\begin{figure}
\begin{center}
\includegraphics[width=\columnwidth]{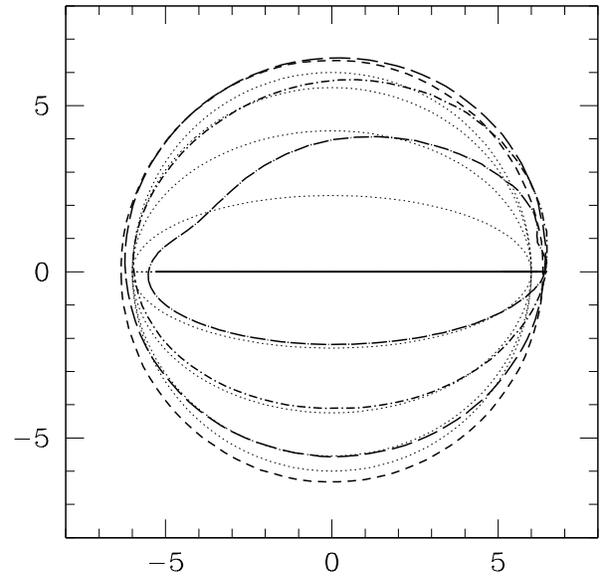} 
\end{center}
\caption{Motion of the centroid of the intensity for circular orbits around
a Schwarzschild black hole with radius $6M$ viewed from $0^\circ$ (solid),
$22.5^\circ$ (long dash-dot), $45^\circ$ (short dash-dot), $67.5^\circ$
(long dash), and $89^\circ$ (short dash) above the orbital plane.  For
reference, circles inclined at the various angles are also shown by the
dotted lines. Axes are labelled in units of $M$.}
\label{theta_cents}
\end{figure}

\begin{figure}
\begin{center}
\includegraphics[width=\columnwidth]{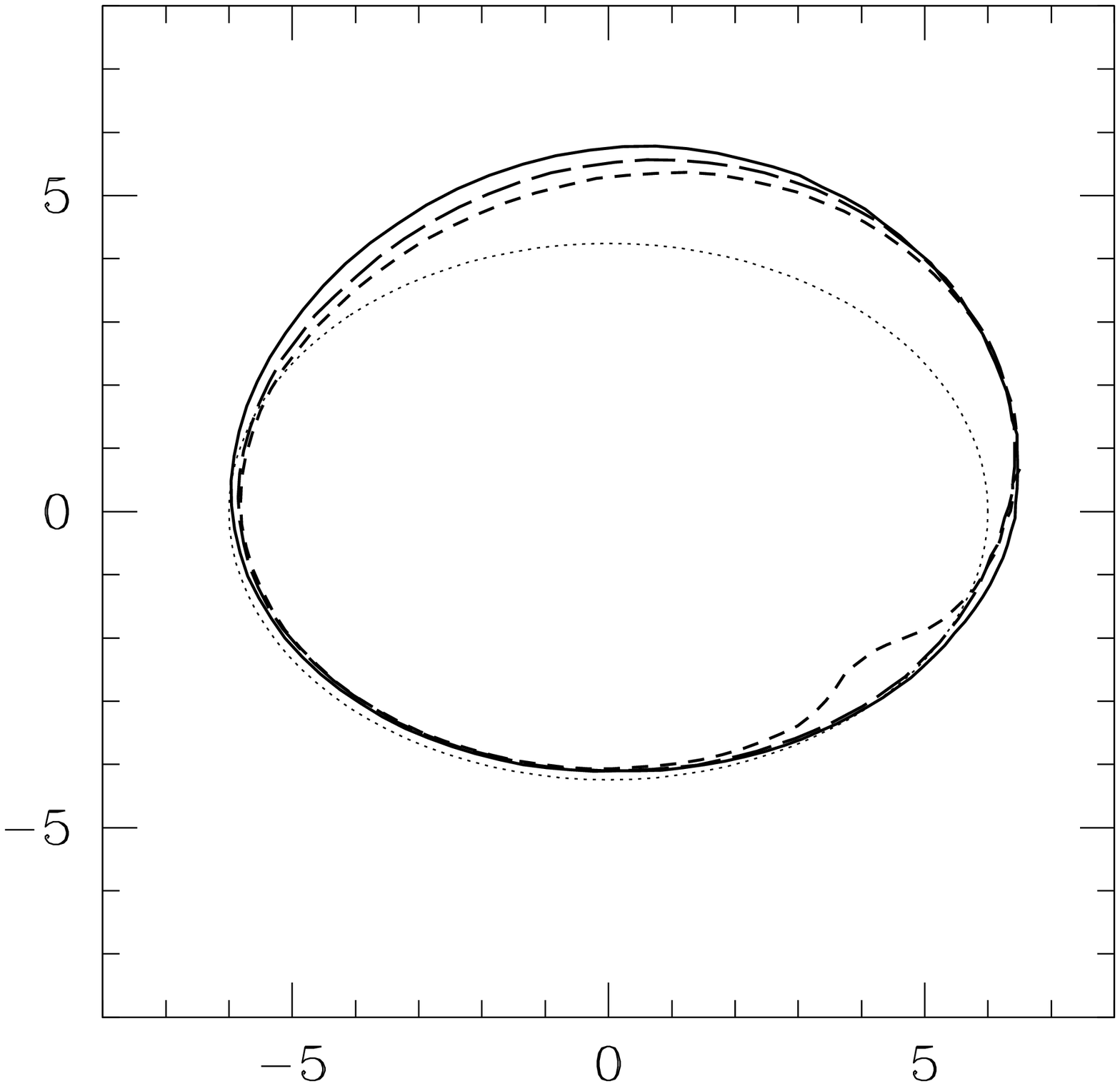} 
\end{center}
\caption{Motion of the centroid of the intensity for circular orbits in the
equatorial plane around a Kerr black hole with radius $6M$ viewed from
$45^\circ$ orbital plane for Kerr spin-parameters $0$, $0.5$, and $0.998$.
For reference, a circle inclined at $45^\circ$ is also shown by the
dotted line.  Axes are labelled in units of $M$.}
\label{a_cents}
\end{figure}

\begin{figure}
\begin{center}
\includegraphics[width=\columnwidth]{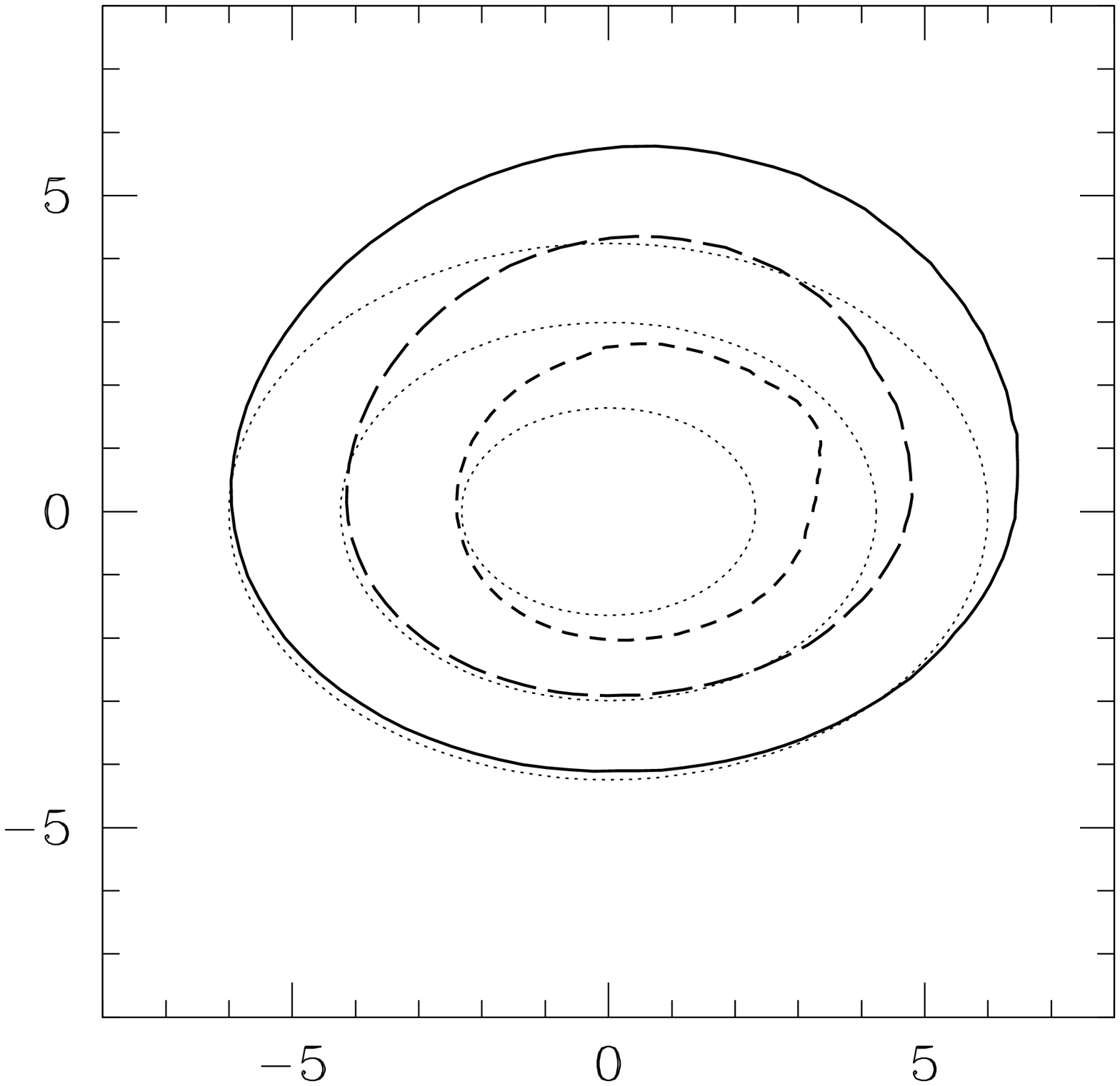} 
\end{center}
\caption{Motion of the centroid of the intensity for the prograde ISCOs
around a Kerr black hole viewed from $45^\circ$ above the orbital plane
with Kerr spin-parameters $0$ (solid), $0.5$ (long dash), and $0.9$ (short
dash).  For reference, circles inclined at $45^\circ$ at the appropriate
radii are also shown by the dotted lines.  Axes are labelled in units of
$M$.}
\label{pisco_cents}
\end{figure}

Figures \ref{r_cents}-\ref{pisco_cents} compare the nearly elliptical
centroid paths for various orbital and black hole parameters.  The
major-axis of the elliptical path is indicative of the orbital radius in a similar
way as it is for Newtonian orbits (Figure \ref{r_cents}).  In general,
gravitational lensing will increase the minor-axis.  Nevertheless, the
minor-axis may be used to infer the orbital inclination (Figure
\ref{theta_cents}).  In contrast with variations of the orbital radii,
gravitational lensing results in changes that are substantially different
from the Newtonian case.  Nonetheless, these results suggest that it
is indeed possible to place significant constraints upon the orbital
parameters from the centroid motion alone.

Of more interest is the possibility of measuring the black hole spin.  Given the
orbital radius and its period, equation (\ref{period_eq}) may be used to
deduce the black hole spin.  Figures \ref{a_cents} and \ref{pisco_cents}
show that this may be constrained by the centroid paths as well.  However,
except for a feature present only near the maximally rotating case,
the paths are very similar, and thus it is likely that imaging will be required to
distinguish the different cases.  Nevertheless, it is not clear that this
is the appropriate comparison because the prograde ISCO moves substantially
inwards with increasing black hole spin.  Therefore, it may be more
suitable to compare different black hole spins at their respective prograde ISCOs, as
shown in Figure \ref{pisco_cents}.  In this case, there is a morphological
change in the centroid paths, in addition to the difference that results
trivially from the decrease in orbital radius.  As a result, when
combined with period measurements, an accurate determination of the
centroid path provides a method with which to measure $a$.

\begin{figure}
\begin{center}
\includegraphics[width=\columnwidth]{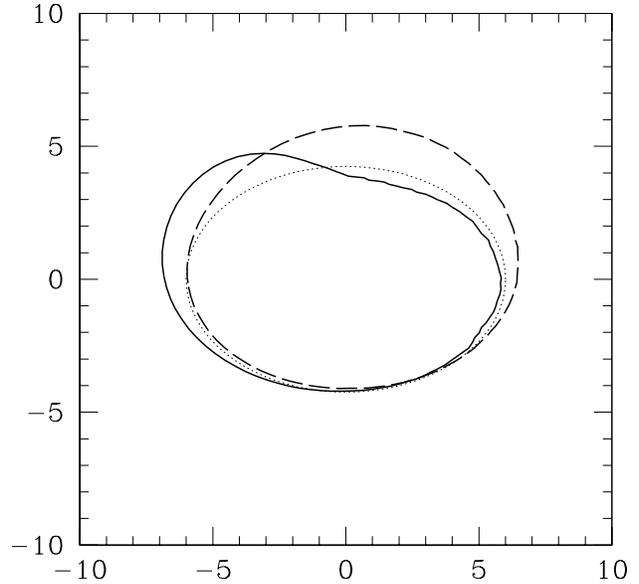} 
\end{center}
\caption{Motion of the centroid of the intensity for circular orbits around
a Schwarzschild black hole with radius $6M$ viewed from $45^\circ$
above the orbital plane for emission models with spectral index
$1.3$ (solid), as in Figure \ref{p6_a0_th45_avg_synch}, and $-2$
(dashed), as in the Rayleigh-Jeans approximation.  For
reference, a circle inclined at $45^\circ$ is also shown by the
dotted lines. Axes are labelled in units of $M$.}
\label{nu_cents}
\end{figure}

As suggested by Figure \ref{p6_a0_th45_avg_synch}, for emission models
with spectral index $>-2$ the centroid position will be dominated by
the portion of the orbit in which the bright spot is moving towards
the observer.  This is shown in Figure \ref{nu_cents}, in which the
centroid paths of the Rayliegh-Jeans and $\alpha=1.3$ emission models
are compared.  Thus, estimates of the orbital parameters from the
motion of the centroid may require additional information about the
spectrum of the emission.  However, as mentioned before, even in the
absence of spectroscopy, the spectral index may be obtained from brightness
asymmetry in the the time-averaged images (see, \eg, equation
\ref{asym_bright}).

\section{Conclusions} \label{C}
The orbit-integrated image of a bright sphere (hot spot)
in motion near the horizon of \SgrA~is qualitatively different from the
image produced by a quasi-spherical accretion flow.  In particular, the
time-averaged image of a compact hot spot does not reveal a clear black hole shadow,
as found for spherical accretion flows \citep{Falc-Meli-Agol:00}.  

The evidence for short-term flaring activity in the Galactic centre
\citep{Ghez_etal:04,Ecka_etal:04,Genz_etal:03,Porq_etal:03,Asch_etal:04,Gold_etal:03,Baga_etal:01}, implies
that the accretion flow around the
horizon of \SgrA~is clumpy or unsteady. In the likely event that the
accretion flow has bright spots, imaging of these
spots could be utilised as a method for inferring the black hole
parameters.  In particular, the signature of the secondary Einstein ring/arc in
the light curves and the images could potentially be more sensitive to
strong gravity effects than imaging the black hole shadow alone.  In the time
averaged images, black hole spin produces relative offsets between features
(see, Figure \ref{pISCO_a09_th45_avg}), which are amenable to interferometric
imaging in a way that measurements of the shadow are not.

Imaging a bright spot may also independently constrain the
mass and distance of \SgrA.  Timing data combined with the projection
of the orbit on the sky (or the apparent velocity) allows the measurement of
the ratio of \SgrA's mass and distance.  If the line-of-sight velocity
of the inhomogeneity can be determined as well (\eg, from spectral
measurements combined with the magnitude of the brightness asymmetry),
this degeneracy can be broken.

If the emission is polarised (and the nature of the intrinsic polarisation
is understood), light curves and images of the polarisation are strongly
dependent upon those features in the image that are produced near the
photon orbit.  Hence, polarisation in general may be diagnostic of the
black hole parameters.

Lastly, even if high precision images are unavailable, measurements of
the centroid path coupled with observations of the light curve provide
an alternative method by which to determine the black hole spin.

In reality, images of the Galactic centre may be expected to result from
inhomogeneities during flaring events superimposed upon a homogeneous
background.  Thus, the results of this paper are complimentary to 
previous work.  It is therefore likely that observations of both, a black
hole shadow between flaring events and the dynamical and/or averaged
properties of the images during flaring events, can be coupled to reduce the
uncertainty inherent to each.

\section*{Acknowledgements}
This work was supported in part by NASA grant NAG 5-13292, and by NSF
grants AST-0071019, AST-0204514 (for A.L.).  A.E.B. gratefully
acknowledges the support of an ITC Fellowship from Harvard College
Observatory.

\bibliographystyle{mn2e.bst}
\bibliography{blobs.bib}

\bsp

\end{document}

%% file: standard_abbr.tex
\def\cf{{cf.}} 
\def\eg{{e.g.}} 
\def\ie{{i.e.}} 
\def\etc{{etc}} 

%% file: unit_abbr.tex
\def\s{{\rm s}} 

\def\pc{{\rm pc}} 
\def\kpc{{\rm k}\pc} 

\def\muas{\mu{\rm as}} 